\documentclass[11pt]{article}
\usepackage[margin=.9in,left=.9in]{geometry}
\usepackage{amsmath}
\usepackage{amsfonts}
\usepackage{amssymb}
\usepackage{amsthm}
\usepackage{mathtools}
\usepackage{subcaption}
\usepackage{graphicx}
\usepackage{color}
\usepackage{xcolor}
\usepackage{xfrac}
\usepackage{stmaryrd}

\usepackage[safe]{tipa} 
\usepackage[utf8]{inputenc}
\usepackage{rotating}
\usepackage{hyperref}
\usepackage[all]{xy}

\usepackage{mathtools}
\usepackage{rotating}

\usepackage{stmaryrd}
\usepackage{multirow}

\usepackage{lmodern} 

\usepackage{tikz}
\usetikzlibrary{decorations.pathmorphing}
\usepackage{tikz-cd}
\usetikzlibrary{calc}
\usetikzlibrary{matrix}
\usetikzlibrary{positioning,arrows,shapes}
\usetikzlibrary{intersections,shapes.arrows, calc}
\usetikzlibrary{patterns}
\usetikzlibrary{mindmap} 
\usetikzlibrary[mindmap] 

\usepackage{lipsum}
\usepackage{adjustbox}

\definecolor{darkblue}{rgb}{0.05,0.25,0.65}
\definecolor{greenii}{RGB}{20,140,10}
\definecolor{lightgray}{rgb}{0.9,0.9,0.9}
\definecolor{orangeii}{RGB}{200,100,5}

\def\acts{\raisebox{1.4pt}{\;\rotatebox[origin=c]{90}{$\curvearrowright$}}\hspace{.5pt}}

\usepackage{multirow}

\usepackage{stmaryrd}    

\usepackage{enumerate} 

\usepackage{helvet}   

\usepackage{bigints}  

\usepackage{scalerel} 

\usepackage{mathptmx}
\usepackage{amsmath}
\usepackage{graphicx}

\usepackage{floatflt}  
\usepackage{array}     
\newcolumntype{L}[1]{>{\raggedright\let\newline\\\arraybackslash\hspace{0pt}}m{#1}}
\newcolumntype{C}[1]{>{\centering\let\newline\\\arraybackslash\hspace{0pt}}m{#1}}
\newcolumntype{R}[1]{>{\raggedleft\let\newline\\\arraybackslash\hspace{0pt}}m{#1}}

\makeatletter
\newcommand{\raisemath}[1]{\mathpalette{\raisem@th{#1}}}
\newcommand{\raisem@th}[3]{\raisebox{#1}{$#2#3$}}
\makeatother

\usepackage{tabularx}   

\newcommand{\BaseSpace}{
  Q
}

\newcommand{\NormalBundle}{
  N Q
}

\newdir{> }{{}*!/10pt/@{>}}

\DeclareRobustCommand{\rchi}{{\mathpalette\irchi\relax}}
\newcommand{\irchi}[2]{\raisebox{\depth}{$#1\chi$}} 

\makeatletter
\newif\if@sup
\newtoks\@sups
\def\append@sup#1{\edef\act{\noexpand\@sups={\the\@sups #1}}\act}%
\def\reset@sup{\@supfalse\@sups={}}%
\def\mk@scripts#1#2{\if #2/ \if@sup ^{\the\@sups}\fi \else%
  \ifx #1_ \if@sup ^{\the\@sups}\reset@sup \fi {}_{#2}%
  \else \append@sup#2 \@suptrue \fi%
  \expandafter\mk@scripts\fi}
\def\tensor#1#2{\reset@sup#1\mk@scripts#2_/}
\def\multiscripts#1#2#3{\reset@sup{}\mk@scripts#1_/#2%
  \reset@sup\mk@scripts#3_/}
\makeatother

\makeatletter
\newbox\slashbox \setbox\slashbox=\hbox{$/$}
\def\itex@pslash#1{\setbox\@tempboxa=\hbox{$#1$}
  \@tempdima=0.5\wd\slashbox \advance\@tempdima 0.5\wd\@tempboxa
  \copy\slashbox \kern-\@tempdima \box\@tempboxa}
\def\slash{\protect\itex@pslash}
\makeatother

\def\clap#1{\hbox to 0pt{\hss#1\hss}}
\def\mathllap{\mathpalette\mathllapinternal}
\def\mathrlap{\mathpalette\mathrlapinternal}
\def\mathclap{\mathpalette\mathclapinternal}
\def\mathllapinternal#1#2{\llap{$\mathsurround=0pt#1{#2}$}}
\def\mathrlapinternal#1#2{\rlap{$\mathsurround=0pt#1{#2}$}}
\def\mathclapinternal#1#2{\clap{$\mathsurround=0pt#1{#2}$}}

\let\oldroot\root
\def\root#1#2{\oldroot #1 \of{#2}}
\renewcommand{\sqrt}[2][]{\oldroot #1 \of{#2}}

\DeclareSymbolFont{symbolsC}{U}{txsyc}{m}{n}
\SetSymbolFont{symbolsC}{bold}{U}{txsyc}{bx}{n}
\DeclareFontSubstitution{U}{txsyc}{m}{n}

\DeclareSymbolFont{stmry}{U}{stmry}{m}{n}
\SetSymbolFont{stmry}{bold}{U}{stmry}{b}{n}

\DeclareFontFamily{OMX}{MnSymbolE}{}
\DeclareSymbolFont{mnomx}{OMX}{MnSymbolE}{m}{n}
\SetSymbolFont{mnomx}{bold}{OMX}{MnSymbolE}{b}{n}
\DeclareFontShape{OMX}{MnSymbolE}{m}{n}{
    <-6>  MnSymbolE5
   <6-7>  MnSymbolE6
   <7-8>  MnSymbolE7
   <8-9>  MnSymbolE8
   <9-10> MnSymbolE9
  <10-12> MnSymbolE10
  <12->   MnSymbolE12}{}

\makeatletter
\def\Decl@Mn@Delim#1#2#3#4{%
  \if\relax\noexpand#1%
    \let#1\undefined
  \fi
  \DeclareMathDelimiter{#1}{#2}{#3}{#4}{#3}{#4}}
\def\Decl@Mn@Open#1#2#3{\Decl@Mn@Delim{#1}{\mathopen}{#2}{#3}}
\def\Decl@Mn@Close#1#2#3{\Decl@Mn@Delim{#1}{\mathclose}{#2}{#3}}
\Decl@Mn@Open{\llangle}{mnomx}{'164}
\Decl@Mn@Close{\rrangle}{mnomx}{'171}
\Decl@Mn@Open{\lmoustache}{mnomx}{'245}
\Decl@Mn@Close{\rmoustache}{mnomx}{'244}
\makeatother

\makeatletter
\DeclareRobustCommand\widecheck[1]{{\mathpalette\@widecheck{#1}}}
\def\@widecheck#1#2{%
    \setbox\z@\hbox{\m@th$#1#2$}%
    \setbox\tw@\hbox{\m@th$#1%
       \widehat{%
          \vrule\@width\z@\@height\ht\z@
          \vrule\@height\z@\@width\wd\z@}$}%
    \dp\tw@-\ht\z@
    \@tempdima\ht\z@ \advance\@tempdima2\ht\tw@ \divide\@tempdima\thr@@
    \setbox\tw@\hbox{%
       \raise\@tempdima\hbox{\scalebox{1}[-1]{\lower\@tempdima\box
\tw@}}}%
    {\ooalign{\box\tw@ \cr \box\z@}}}
\makeatother

\makeatletter
\def\udots{\mathinner{\mkern2mu\raise\p@\hbox{.}
\mkern2mu\raise4\p@\hbox{.}\mkern1mu
\raise7\p@\vbox{\kern7\p@\hbox{.}}\mkern1mu}}
\makeatother

\def\1{{\bf 1}}

\def\<{\langle}
\def\>{\rangle}

\renewcommand{\(}{\begin{equation}}
\renewcommand{\)}{\end{equation}}
\newcommand{\bea}{\begin{eqnarray*}}
\newcommand{\eea}{\end{eqnarray*}}

\usepackage{cleveref}

\crefformat{section}{\S#2#1#3} 
\crefformat{subsection}{\S#2#1#3}
\crefformat{subsubsection}{\S#2#1#3}

\theoremstyle{italics}
\newtheorem{theorem}{Theorem}

\newtheorem{prop}[theorem]{Proposition}
\newtheorem{cor}[theorem]{Corollary}

\theoremstyle{definition}
\newtheorem{defn}[theorem]{Definition}

\newtheorem{example}[theorem]{Example}

\newtheorem{remark}[theorem]{Remark}
\newtheorem{note[theorem]}{Note}

\usepackage{amsfonts}
\usepackage{colortbl}

\definecolor{darkblue}{rgb}{0.05,0.25,0.65}
\definecolor{darkgreen}{rgb}{0.00,0.85,0.1}
\definecolor{plum}{rgb}{0.36078, 0.20784, 0.4}

\usepackage{hyperref}

\begin{document}

\title{Twisted Cohomotopy implies M5-brane anomaly cancellation}

\author{
  Hisham Sati\thanks{\tt hsati@nyu.edu},
  \quad
  Urs Schreiber\thanks{{\tt us13@nyu.edu} (corresponding author) \newline}
}

\maketitle

\begin{abstract}
We highlight what seems to be a remaining subtlety in the argument for the cancellation
of the total anomaly associated with the M5-brane in M-theory. Then we prove that this
subtlety is resolved under the hypothesis that the C-field flux is charge-quantized
in the generalized cohomology theory called J-twisted Cohomotopy.
\end{abstract}

\tableofcontents

\section{Introduction}

Formulating M-theory remains an open problem
(e.g. \cite[6]{Duff96}\cite[p. 2]{HoweLambertWest97}\cite[p. 6]{Duff98}\cite[p. 2]{NicolaiHelling98}\cite[p. 330]{Duff99}\cite[12]{Moore14}\cite[p. 2]{ChesterPerlmutter18}\cite[@21:15]{Witten19}\cite[@17:04]{Duff19}).
Even formulating just the field-theoretic decoupling limit of the worldvolume theory
of M5-branes in M-theory remains an open problem (e.g. \cite[6.3]{Lambert19}).
Nevertheless, it is traditionally assumed that enough is known about
M-theory in general, and about M5-branes in particular, that
it makes sense to check whether field theoretic anomalies
(following \cite{AW84}\cite{AG85}) on M5-brane worldvolumes
cancel  against M-theoretic anomaly inflow (following \cite{CallanHarvey85})
from the bulk spacetime (reviewed in the current context in \cite{Harvey05}).

\medskip

\noindent {\bf Relevance of anomaly cancellation for M-theory.}
What from the physics perspective are called \emph{anomalies}
is what from the perspective of mathematics are \emph{obstructions}
(a point highlighted in \cite{KS1}\cite{SSS2}).
Hence such a cancellation of the total M5-brane anomaly,
if properly identified, is strictly necessary for M-theory to exist: any remaining anomaly
is an obstruction against the existence of the theory of which it
is an anomaly. But conversely, wherever a putative anomaly in M-theory
is found \emph{not} to vanish, by available reasoning,
this signifies (with the assumption that M-theory does in fact exist)
the presence of a new aspect of the elusive theory
that had hitherto been missed: There must then be a new
detail in the theory, previously unrecognized,
which does imply the cancellation of the remaining anomaly, after all.

\medskip
For this reason a careful mathematical analysis of anomaly cancellation
in M-theory is in order. The tacit assumption that the proverbial magic of M-theory
will take care of all cancellations anyway, freeing us from the
burden of patient rigorous checks, would work only if the actual formulation of M-theory
were known. Since it is not known,
the situation is the reverse: A carefully deduced failure of anomalies
to cancel provides a hint as to the actual formulation of the elusive theory.

\medskip

\noindent {\bf Historical background on M5-brane anomaly cancellation.}
Indeed, the original computation of the total M5-brane anomaly in
\cite[5]{Witten96} found the total anomaly \emph{not} to vanish;
and highlighted that the issue remains an open problem
(``somewhat puzzling'' \cite[p. 35]{Witten96}).
In reaction, several authors argued for several fixes,
but, it seems, without
convincing success (see \cite[p.2]{FHMM98} for pointers).
Finally, \cite[3]{FHMM98} argued that there is a previously neglected summand
in the bulk anomaly inflow which needs to be taken into account
(the top right term in diagram \eqref{TheAnomalyPolynomials} below).
That correction to the bulk anomaly inflow term
has since become accepted (e.g., in \cite[(5)]{BahBonettiMinasianNardoni18})
as the solution to the M5-brane anomaly cancellation.
The authors of \cite[A.4-5]{BahBonettiMinasianNardoni19}
recently recall the argument of \cite{FHMM98} in streamlined form.
Nonetheless, these arguments remain non-rigorous
even by physics standards, due to a lack of actual formulation of M-theory.
This is clearly acknowledged and highlighted by one of these authors,
in \cite[p. 46]{Harvey05}.\footnote{
\cite[p. 46]{Harvey05}: ``[...] the solution is not so clear. [The established procedure of anomaly cancellation]
will not work for the M5-brane. [...] something new is required. What this something new is, is not a priori
obvious. [...] [This is] a daunting task. To my knowledge no serious attempts have been made to study the
problem. [...] [The proposal of \cite{FHMM98}]  probably should not be viewed as a final understanding
of the problem. One would eventually hope for a microscopic formulation of M-theory which makes some
of the manipulations [proposed in \cite{FHMM98}] appear more natural.''}

\medskip
\noindent {\bf Remaining issue.}
In this note we point out, in \cref{TheIssue} below,
that there does still remain one issue
with the currently accepted anomaly cancellation argument
\cite[3]{FHMM98}\cite[A.4-5]{BahBonettiMinasianNardoni19} in itself.
This is a simple observation: these authors made an
\emph{Ansatz} (see \eqref{TheAnsatz} below) for the C-field configuration
(\cite[(2.3)]{FHMM98}\cite[(A.18)]{BahBonettiMinasianNardoni19})
which is seemingly not the most general admissible
under the given assumptions (as also noticed in \cite[(3.12)]{Monnier13}\cite[(3.16)]{BahBonettiMinasianNardoni19b}\cite[(2.34)]{BahBonettiMinasian20}).
Entering their anomaly cancellation
argument instead with a general C-field configuration
seemingly leaves one anomaly contribution uncancelled,
shown on the bottom right of  \eqref{TheAnomalyPolynomials} below.

\medskip

\noindent {\bf Resolution by Hypothesis H.}
We prove in \cref{TheSolution} that this previously
neglected remaining anomaly term does in fact vanish, hence that the
anomaly cancellation argument of
\cite[\S 5]{Witten96}\cite[\S 3]{FHMM98}\cite[A.4-5]{BahBonettiMinasianNardoni19}
is completed, if one assumes
a hypothesis \cite[\S 2.5]{Sati13}
about the proper nature of the C-field in
M-theory  which in \cite{FSS19b}\cite{FSS19c}\cite{SS19a}\cite{SS19c}\cite{SS21}
we called \hyperlink{HypothesisH}{\it Hypothesis H},
recalled in \cref{ViaHypothesisH} below.
This hypothesis says that the M-theory C-field
is charge-quantized \eqref{GeneralTwistedCharacterMap}
in the generalized cohomology theory
called \emph{J-twisted Cohomotopy} \eqref{TwistedCohomotopicalCharacter}.
We have previously demonstrated
that this hypothesis implies a wealth of further anomaly cancellation
conditions \cite{FSS19b}\cite{FSS19c}\cite{SS19a}\cite{SS19b}
and other effects \cite{SS19c}\cite{SS21} expected in M-theory (exposition in \cite{Sc20}).

\medskip

\noindent {\bf Outlook.}
Since \hyperlink{HypothesisH}{\it Hypothesis H} gives rigorous
mathematical meaning to the M-theoretic nature of the C-field,
our derivation in \cref{TheSolution} is a rigorous mathematical
proof of the vanishing of the remaining anomaly term \eqref{TheAnomalyPolynomials}
from this hypothesis and, as such, completes the argument
of \cite[\S 5]{Witten96}\cite[\S 3]{FHMM98}\cite[A.4-5]{BahBonettiMinasianNardoni19}.
We do not claim to make the rest of that argument rigorous.
In order to do so one will need also
a rigorous definition of the M5-brane coupled to this C-field.
We have presented results going towards that goal in
\cite{FSS19d}\cite{FSS20}\cite{FSS20a}\cite{FSS20b}\cite{FSS19c},
but more needs to be done.

\medskip
\medskip

\noindent {\bf Acknowledgement.}
We thank Domenico Fiorenza for very useful discussions.
We thank an anonymous referee for alerting us of \cite{BahBonettiMinasianNardoni19b},
whose Section 4.1 overlaps with the
discussion in our \cref{ViaGenericTadpoleCancellation}; see
Remark \ref{ComparisonToBBMN19b}.

\newpage

\section{The issue}
\label{TheIssue}

\noindent {\bf The geometry under consideration.} We are dealing with
(for background see e.g. \cite{Duff99}\cite{MiemicSchnakenburg06}):

\noindent \hspace{-.1cm}
\begin{tabular}{ll}
 {\bf (i)} & families of
 \\
 {\bf (ii)} & C-field configurations on
 \\
 {\bf (iii)} & 11-dimensional spacetimes
 \\
 {\bf (iv)} & sourced by magnetic 5-branes
 \\
 {\bf (v)} & of unit charge.
\end{tabular}

\vspace{.1cm}

\noindent We now say what this means precisely: First, {\bf (i)} with {\bf (iii)} means that
$$
  X
   \;:=\;
  \overset{
    \mathclap{
    \mbox{
      \tiny
 \bf     \color{darkblue}
      \begin{tabular}{c}
        spacetime
        \\
        manifold
      \end{tabular}
    }
    }
  }{
    \overbrace{
      X^{11}
    }
  }
  \;\;\times\;\;
  \overset{
    \mathclap{
    \mbox{
      \tiny
\bf      \color{darkblue}
      \begin{tabular}{c}
        parameter
        \\
        manifold
      \end{tabular}
    }
    }
  }{
    \overbrace{
      U
    }
  }
$$
is the product of an 11-dimensional
manifold (spacetime) with a parameter manifold $U$ of any dimension,
while {\bf (ii)} means that we consider a closed differential 4-form
on $X$:
$$
  \underset{
    \mathclap{
    \mbox{
      \tiny
 \bf     \color{darkblue}
      \begin{tabular}{c}
        family of
        \\
        C-field flux densities
      \end{tabular}
    }
    }
  }{
    \underbrace{
      G_4
    }
  }
    \in
  \Omega^4_{\mathrm{cl}}(X)
  \;\;\;
  \Longrightarrow
  \;\;\;
  \underset{s \in U}{\forall}
  \Big(\;\;\;
    \underset{
      \mathclap{
      \mbox{
        \tiny
 \bf       \color{darkblue}
        \begin{tabular}{c}
          C-field flux density
          \\
          at parameter $s$
        \end{tabular}
      }
      }
    }{
      \underbrace{
        G_4^{(s)}
      }
    }
    \;\in\;
    \Omega^4_{\mathrm{cl}}(X^{11})
  \Big)
  \,,
$$
which hence is, in particular,
a $U$-parametrized
family of differential 4-forms on $X^{11}$.\footnote{The inclined reader may think of the 4-flux data
$G_4$ as being a value at stage
$U$ of the \emph{mapping stack}
$\mathbf{Fields}(X^{11}) := [X^{11},\underline{\Omega}^4]$
into the sheaf of differential 4-forms,
and of the anomaly polynomials \eqref{TheAnomalyPolynomials} as
being (classes of) differential forms
on this mapping stack.
While this is the correct point of view (exposition in \cite{FSS13}),
we will not further dwell on it here.}
Moreover, {\bf (iv)} means, just as in Dirac's argument for
magnetic 0-branes (e.g. \cite[\S 2]{Alvarez85}), that $X^{11}$ is the complement of a 5-brane
worldvolume, hence that $X$ is an orthogonal $S^4$-fiber bundle
(see Def. \ref{OrthonalSphereFiberBundle})
as shown on the left of
\eqref{The4SphericalFibration}.
\begin{equation}
  \label{The4SphericalFibration}
  \hspace{1cm}
  \raisebox{30pt}{
  \xymatrix{
    \mathllap{
      \mbox{
        \tiny
 \bf       \color{darkblue}
        \begin{tabular}{c}
          unit sphere
          \\
          around
          \\
          M5-brane
        \end{tabular}
      }
      \;\;
    }
    S^4
    \ar[r]
    &
    \overset{
      \mathclap{
      \mbox{
        \tiny
        \begin{tabular}{c}
      \bf    \color{darkblue}
          spacetime
          \\
          (families)
          \\
          \phantom{a}
        \end{tabular}
      }
      }
    }{
      X
    }
    \ar[d]^-{
      S(p)
    }_-{
      \mathllap{
      \mbox{
        \tiny
        \begin{tabular}{c}
   \bf       \color{darkblue}
          4-sphere
          \\
          \color{darkblue}
          fiber bundle
        \end{tabular}
      }
      }
    }
    &
    &
    \overset{
      \mathclap{
      \mbox{
        \tiny
        \begin{tabular}{c}
  \bf        \color{darkblue}
          C-field
          \\
    \bf      \color{darkblue}
          4-flux density
          \\
          (in families)
          \\
          \phantom{a}
        \end{tabular}
      }
      }
    }{
      [G_4]
    }
    \ar@{|->}[d]|-{
      \mbox{
        \tiny
 \bf       \color{darkblue}
        \begin{tabular}{c}
          total flux
          \\
          through $S^4$
        \end{tabular}
      }
    }
    \ar@{}[r]|-{\in}
    &
    \overset{
      \mathclap{
      \mbox{
        \tiny
 \bf       \color{darkblue}
        \begin{tabular}{c}
          de Rham
          \\
          cohomology
          \\
          \phantom{a}
        \end{tabular}
      }
      }
    }{
      H^{\bullet + 4}_{\mathrm{dR}}(X)
    }
    \ar[d]_-{ \int_{S^4} }^-{
      \mathrlap{
        \mbox{
          \tiny
          \begin{tabular}{c}
  \bf          \color{darkblue}
            fiber integration
          \end{tabular}
        }
      }
    }
    \ar@{}[r]|-{\simeq}
    &
    \overset{
      \mathclap{
      \mbox{
        \tiny
    \bf    \color{darkblue}
        \begin{tabular}{c}
          real
          \\
          cohomology
          \\
          \phantom{a}
        \end{tabular}
      }
      }
    }{
      H^{\bullet + 4}(X; \mathbb{R})
    }
    \ar[d]^-{
      S(p)_\ast
    }
    \\
    &
    \mathllap{
  \underset{
    \mathclap{
    \mbox{
      \tiny
 \bf     \color{darkblue}
      \begin{tabular}{c}
        \phantom{a}
        \\
        parameter
        \\
        manifold
      \end{tabular}
    }
    }
  }{
    U
  }
  \;\times\;
  \underset{
    \mathclap{
    \mbox{
      \tiny
 \bf     \color{darkblue}
      \begin{tabular}{c}
        $\phantom{a}$
        \\
        radial
        \\
        distance
        \\
        from brane
      \end{tabular}
    }
    }
  }{
    (0,\infty)
  }
   \;\times\;
  \underset{
    \mathclap{
    \mbox{
      \tiny
      \begin{tabular}{c}
        $\phantom{a}$
        \\
  \bf      \color{darkblue}
        M5-brane
        \\
   \bf     \color{darkblue}
        worldvolume
        \\
        (families)
      \end{tabular}
    }
    }
  }{
      Q_{\mathrm{M5}}
  }
    \;
    =
    \;
    }
    \BaseSpace
    &
    &
    \underset{
      \mathclap{
      \mbox{
        \tiny
        \begin{tabular}{c}
          $\phantom{a}$
          \\
   \bf       \color{darkblue}
          single M5
          \\
          ($\Leftrightarrow$ abelian 2-form field)
        \end{tabular}
      }
      }
    }{
      1
    }
    \ar@{}[r]|-{ \in }
    &
    H^\bullet_{\mathrm{dR}}(\BaseSpace)
    \ar@{}[r]|-{\simeq}
    &
    H^\bullet(\BaseSpace; \mathbb{R} )
  }
  }
\end{equation}
Finally, {\bf (v)} means that the corresponding fiber integration
\eqref{The4SphericalFibration}
of $G_4$ over the 4-sphere fibers is unity\footnote{
   Our derivations in \cref{TheSolution} immediately apply generally to
   any integer charge $S(p)_\ast[G_4] \in \mathbb{N}$ \eqref{UnitFluxCondition}.
   But for $N \geq 2$ even the nature of the higher gauge field on the M5-brane(s)
   remains open (see \cite{FSS20} for pointers and for a resolution for $N = 2$)
   and it seems premature to extrapolate the existing computations of worldvolume anomalies
   to this case (compare \cite[below (2.4)]{HarveyMinasianMoore98}).
 }
\begin{equation}
  \label{UnitFluxCondition}
  S(p)_\ast [G_4]
  \;=\;
  1
  \;\in\;
  H^0(\BaseSpace; \mathbb{R})
\end{equation}
as shown on the right of \eqref{The4SphericalFibration}.
The general solution to \eqref{UnitFluxCondition} is
the sum of half the Euler class of the $S^4$-fibration
(e.g. \cite[\S 11]{BottTu82}\cite[(2.3)]{BottCattaneo97})
with any \emph{basic} class
(by exactness of the Gysin sequence, e.g. \cite[14.33]{BottTu82}),
namely one pulled back
from the base of the fibration:

\vspace{-.2cm}

\begin{equation}
  \label{Density4Form}
  \raisebox{53pt}{
  \xymatrix@R=-2pt{
    &&
    \big[
      \mathrm{vol}_{S^4}
    \big]
    &
    & &
    \mathllap{
      \in
      \;\;\;\;
    }
    H^4(S^4)
    \\
    &
    \mathllap{
      \mbox{
        \tiny
  \bf      \color{darkblue}
        \begin{tabular}{c}
          general
          \\
          4-flux density
          \\
          with unit flux
          \\
          through $S^4$
          \\
          $\phantom{a}$
        \end{tabular}
      }
      \;
    }
    \big[
      G_4
    \big]
    \ar@{}[r]|-{=}
    &
    \underset{
      \mathclap{
      \mbox{
        \tiny
 \bf       \color{darkblue}
        \begin{tabular}{c}
          $\phantom{a}$
          \\
          Euler class of
          \\
          $S^4$-fibration
        \end{tabular}
      }
      }
    }{
      \tfrac{1}{2}\rchi_4
    }
    \ar@{|->}[u]
    \ar@{}[r]|-{+}
    &
    \overset{
      \mathclap{
      \mbox{
        \tiny
  \bf      \color{darkblue}
        \begin{tabular}{c}
          basic component:
          \\
          pulled back from
          \\
          base of $S^4$-fibration
          \\
          $\phantom{a}$
        \end{tabular}
      }
      }
    }{
      S(p)^\ast\big[G_4^{{}^{\mathrm{basic}}}\big]
    }
    &&
    \mathllap{
      \in
      \;\;\;\;
    }
    H^4(X)
    \ar[u]^-{ i_x^\ast }
    \\
    &
    &
    &
    \underset{
      \mbox{
        \tiny
    \bf    \color{darkblue}
        \begin{tabular}{c}
          $\phantom{a}$
          \\
          4-class on base
          \\
          of $S^4$-fibration
        \end{tabular}
      }
    }{
      \big[ G_4^{{}^{\mathrm{basic}}} \big]
    }
    \ar@{|->}[u]
    &&
    \mathllap{
      \in
      \;\;\;\;
    }
    H^4(\BaseSpace)
    \ar@/_2.4pc/[uu]_-{0}
    \ar[u]^-{S(p)^\ast}
  }
  }
\end{equation}


\vspace{-.4cm}

\begin{remark}[The $\sfrac{1}{2}$BPS M5 configuration and its generalization]
  The local model of the situation \eqref{The4SphericalFibration} is the
  trivial $S^4$-fibration of the near horizon geometry
  of the smooth $\sfrac{1}{2}$-BPS black M5-brane solution of
  11-dimensional supergravity
  (\cite{GibbonsTownsend93}, reviewed in \cite[\S 2.1.2]{AFHS98}), restricted to the
  Poincar{\'e} patch of 7-dimensional anti de-Sitter spacetime:

  \vspace{-.3cm}

  \begin{equation}
    \label{M5AdS7Solution}
    \xymatrix@R=10pt{
      S^4
      \ar[r] &
      \mathrm{AdS}^{\mathrlap{{}^{\mathrm{Poin}}}}_7 \;\; \times S^4
      \ar[d]^-{ S(p) = \mathrm{pr}_1 }
      &
      G_4 = \mathrm{vol}_{{}_{S^4}}
      \\
      &
      \mathllap{
        \underset{
          \mathclap{
          \mbox{
            \tiny
    \bf        \color{darkblue}
            \begin{tabular}{c}
              $\phantom{a}$
              \\
              M5-brane
              \\
              worldvolume
         \   \end{tabular}
          }
          }
        }{
          \mathbb{R}^{5,1}
        }
        \;\times\;
        \underset{
          \mathclap{
          \mbox{
            \tiny
     \bf       \color{darkblue}
            \begin{tabular}{c}
              $\phantom{a}$
              \\
              radial
              \\
              distance
            \end{tabular}
          }
          }
        }
        {
          (0,\infty)
        }
        \;\;\;\;
        \underset{
          \mathrm{diff}
        }{\simeq}
        \;\;\;\;\;
      }
      \underset{
        \;\;\;\;\;\;\;\;\;\;\;
        \mathclap{
        \mbox{
          \tiny
   \bf       \color{darkblue}
          \begin{tabular}{c}
            Poincar{\'e} chart of
            \\
            anti-de Sitter spacetime
          \end{tabular}
        }
        }
      }{
        \mathrm{AdS}^{\mathrlap{{}^{\mathrm{Poin}}}}_7
      }
    }
  \end{equation}
  So the point of \eqref{The4SphericalFibration} is
  to generalize the situation away from this highly symmetric
  $\sfrac{1}{2}$-BPS configuration \eqref{M5AdS7Solution}
  to more general 5-brane configurations.
  While few to no
  black M5-brane solutions to 11d supergravity
  beyond \eqref{M5AdS7Solution} are known explicitly,
  only their topological structure matters for
  the discussion of anomaly cancellation; and that
  topological structure is (essentially by definition)
  what is expressed by \eqref{The4SphericalFibration}.
\end{remark}

\begin{remark}[$G_4$ is singular on the M5-brane locus]
  Condition \eqref{UnitFluxCondition} implies
  (immediately so by the Poincar{\'e} Lemma, since $G_4$ is closed)
  that the flux density $G_4$ can \emph{not}
  be extended to the locus of the M5-brane itself,
  which is (or would be) at the center $r = 0 \in [0,\infty)$
  of the punctured ball $S^4 \times (0,\infty)$ in \eqref{The4SphericalFibration}.
  Instead
  it must/would have a singularity at $r = 0$,
  as is manifest also from the basic example \eqref{M5AdS7Solution}.
  Parts of the literature gloss over this
  subtlety; and the point made in \cite[p. 4-5]{FHMM98} was to argue that
  this is the source of the missing anomaly cancellation
  of \cite{Witten96}. To handle the singularity mathematically,
  these authors declared\footnote{
    \cite[p. 4]{FHMM98}:
    ``We leave to the future the very interesting question of the relation of this
    approach to that based on a direct study of solutions to supergravity.''
  }
  to multiply $G_4$ by a smooth radial cutoff function,
  thus rendering it no longer closed
  \cite[(2.3), (3.4)]{FHMM98} but, mathematically,
  extendable to the brane locus.
  Luckily, the key computation \cite[(3.3)]{FHMM98},
  recalled in \eqref{TheAnomalyPolynomials} below,
  applies just as well if instead one leaves $G_4$ intact
  but removes the singular locus from spacetime,
  just as usual in supergravity \eqref{M5AdS7Solution}.
\end{remark}


\begin{remark}[Focus on real cohomology]
We focus here entirely on the anomaly polynomials
in real cohomology, hence ignoring all torsion contributions
(which become visible in integral cohomology)
as well as all ``global'' anomaly contributions
(which become visible in differential cohomology,
see \cite{FSS20c} for discussion of all these notions of cohomology and their
relations).
Because, while vanishing of the anomaly in
real cohomology is not sufficient for full anomaly cancellation
(which must happen in differential integral cohomology),
it is the \emph{necessary} first step.
No argument about torsion of global contributions to the M5 anomaly
(which, of course, one will eventually want to address)
can affect the proof of anomaly cancellation at the rational/real
approximation; and as long as subtleties do remain here, it behooves
us to first focus on these.
Therefore, we often abbreviate $H^\bullet(-) := H^\bullet(-;\mathbb{R})$,
here and in the following.
\end{remark}

\newpage

\noindent {\bf The anomaly polynomials.}
The cohomology classes contributing to the total
M5-brane anomaly in the situation \eqref{The4SphericalFibration}
are given in the literature as follows:

\vspace{-.2cm}

\begin{equation}
  \label{TheAnomalyPolynomials}
  \raisebox{58pt}{
  \hspace{-1cm}
  \xymatrix@R=-4pt@C=10pt{
    \overset{
      \mbox{
        \tiny
   \bf     \color{darkblue}
        \begin{tabular}{c}
          Bulk
          \\
          spacetime
          \\
          CS-terms
          \\
          $\phantom{a}$
        \end{tabular}
      }
    }{
      H^{12}(X)
    }
    \ar[dd]_-{ S(p)_\ast}^-{
      \!\!\!
      \mathrlap{
      \mbox{
        \tiny
   \bf     \color{darkblue}
        \begin{tabular}{c}
          anomaly
          \\
          inflow
        \end{tabular}
      }
      }
    }
    &
    &
    &
    &
    \big[G_4 \wedge I_8\big]
    \ar@{|->}[dd]
    \ar@{}[rr]|<<<<<<<<{+}
    &
    &
    \mathclap{
      \;\;\;\;\;
      \tfrac{-1}{6} \big[G_4\wedge G_4 \wedge G_4\big]
    }
    \phantom{AAA}
    \ar@{|->}[ddl]
    \ar@{|->}[ddr]
    \\
    {\phantom{A}}
    \\
    \underset{
      \mbox{
        \tiny
  \bf      \color{darkblue}
        \begin{tabular}{c}
          $\phantom{a}$
          \\
          M5-brane
          \\
          worldvolume
          \\
          anomalies
        \end{tabular}
      }
    }{
      H^8(\BaseSpace)
    }
    &
    A_{\scalebox{.5}{$\mathrm{total}$}}
    \;\;\;
    \ar@{}[r]|<{=}
    &
    A_{
      \!\!\!
      \scalebox{.65}{
        $\mathrm{chiral} \atop \mathrm{fermion}$
      }
    }
    \ar@{}[r]|-{+}
    &
    \underset{
      \mathclap{
        \underbrace{
          \phantom{-------------}
        }
      }
    }{
      A_{
        \!
        \scalebox{.65}{
          $\mathrm{chiral} \atop \mathrm{2form}$
        }
      }
    }
    \ar@{}[r]|-{+}
    &
    I_8
    \ar@{}[r]|-{+}
    &
    \tfrac{-1}{24}p_2(\NormalBundle)
    \ar@{}[rr]|-{+}
    &
    &
    \tfrac{-1}{2}
      \big[
      G_4^{{}^{\mathrm{basic}}}
      \wedge
      G_4^{{}^{\mathrm{basic}}}
      \big]
    \ar@{=}[dddd]
    \\
    && &
      \mathllap{
        \scalebox{.8}{
          \cite{Witten96}:
        }
        \;\;
      }
      \tfrac{1}{24}p_2(\NormalBundle)
      &
    \underset{
      \mathclap{
        \;\;
        \underbrace{
          \phantom{-----------------}
        }
      }
    }{
    \phantom{A}
    }
    &
    &&
    \\
    \\
    \\
    &
    & &&
    \mathllap{
      \scalebox{.8}{
        \cite{FHMM98}:
        \;\;
      }
    }
    0
    \ar@{}[rrr]|>>>>>>>>>{+}
    & &
    \underset{
      \!\!\!\!\!\!\!
      \!\!\!\!\!
      \mathclap{
        \underbrace{
          \phantom{----------------------}
        }
      }
    }{
    \phantom{A}
    }
    &
    \tfrac{-1}{2}
      \big[
      G_4^{{}^{\mathrm{basic}}}
      \wedge
      G_4^{{}^{\mathrm{basic}}}
      \big]
    \\
    \mbox{\vspace{-4cm}}
    \\
    {\phantom{a}}
    \\
    & & & & & &
    \!\!\!\!\!\!\!\!\!\!\!\!\;\;\;\;\;\;\;
    \mathllap{
      \scalebox{.8}{
        \hyperlink{HypothesisH}{Hypothesis H}:
        \;\;
      }
    }
    0
  }
  }
\end{equation}

\vspace{2mm}
\noindent We discuss the various items in \eqref{TheAnomalyPolynomials}:

\begin{enumerate}[{\bf (i)}]
\item
The term $I_8$ \eqref{TheI8Term} is the ``one-loop polynomial''
\cite{DuffLiuMinasian95}\cite{VafaWitten95},
while the terms
$A_{\!\!\!\scalebox{.65}{$\mathrm{chiral} \atop \mathrm{fermion}$}}$
and
$A_{\!\scalebox{.65}{$\mathrm{chiral} \atop \mathrm{2form}$}}$
are the plain anomalies \cite[(5.1), (5.4)]{Witten96}
of the chiral fermion and of the abelian chiral
(i.e., with self-dual curvature)
2-form field in 6d QFT. These were expected in \cite{Witten96}
to cancel against the influx of $I_8$, but found there (\cite[(5.7)]{Witten96})
to cancel only up to a remaining term $\tfrac{1}{24}p_2(\NormalBundle)$,
where $\NormalBundle$ denotes the normal
bundle to the M5-brane locus in spacetime.

\item The Chern-Simons term $-\tfrac{1}{6}G_4 \wedge G_4 \wedge G_4$
of 11-dimensional supergravity was argued in \cite[\S 3]{FHMM98}
\cite[A.4-5]{BahBonettiMinasianNardoni19}
to contribute to the anomaly influx from the bulk.
Then a formula due to \cite[Lem 2.1]{BottCattaneo97}
shows that this gives rise to the previously
missing summand of $\tfrac{-1}{24}p_2(\NormalBundle)$.
However, these authors consider an Ansatz for the C-field
configuration
\cite[(2.3), (3.4)]{FHMM98}\cite[(2.4)]{BahBonettiMinasianNardoni19}
which amounts to assuming
\begin{equation}
  \label{TheAnsatz}
  [G_4^{{}^{\mathrm{basic}}}] \overset{!}{=} 0
\end{equation}
in \eqref{Density4Form}.
If this assumption is not made, then the bulk Chern-Simons term
in addition contributes an influx term
$\tfrac{-1}{2}\big[ G_4^{{}^{\mathrm{basic}}}\wedge G_4^{{}^{\mathrm{basic}}}\big]$
(bottom right of \eqref{TheAnomalyPolynomials}),
whose vanishing remains to be discussed.

\item That the Ansatz \eqref{TheAnsatz} remained unjustified
was pointed out in
\cite[(19)]{FSS19a}
and then in \cite[(3.16) \& App. C]{BahBonettiMinasianNardoni19b}
(where the basic component is denoted $\gamma_4$, see also \cite[(2.34)]{BahBonettiMinasian20}).
Previously in \cite[(3.12)]{Monnier13} the term $G_4^{\mathrm{basic}}$
was assumed to be non-vanishing, in general, and
as a resolution it is was suggested \cite[(3.7)]{Monnier13} that
the traditional expression from \cite[(5.7)]{Witten96}
for the self-dual field anomaly
$A_{\!\scalebox{.65}{$\mathrm{chiral} \atop \mathrm{2form}$}}$
in real cohomology
is wrong, in that it gets corrected by just the
seemingly missing summand
$\tfrac{+1}{2}\big[ G_4^{{}^{\mathrm{basic}}}\wedge G_4^{{}^{\mathrm{basic}}}\big]$.
Unfortunately, we are unable to verify this derivation.
Fortunately, it makes no difference:

\item We prove in \cref{TheSolution} that
the seemingly restrictive Ansatz \eqref{TheAnsatz} is \emph{implied}
as soon as the dual $G_7$-flux satisfies a Bianchi identity
of a widely expected form
(Theorem \ref{BasicComponentVanishes} below);
in particular, if it satisfies the Bianchi identity that is implied by
\hyperlink{HypothewsisH}{Hypothesis H}
(Thm. \ref{AnomalyCancellationByHypothesisH} below).
In this way, \hyperlink{HypothewsisH}{Hypothesis H} enforces
the vanishing of the problematic remaining anomaly term by itself:
$$
  \mbox{
    \hyperlink{HypothesisH}{Hypothesis H}
  }
  \;\;\;\;\;\;\;\;
  \Rightarrow
  \;\;\;\;\;\;\;\;
  \big[ G_4^{{}^{\mathrm{basic}}}\big]
  \,=\,
  0
  \;\;\;\;\;\;\;\;
  \Rightarrow
  \;\;\;\;\;\;\;\;
  \big[ G_4^{{}^{\mathrm{basic}}} \wedge G_4^{{}^{\mathrm{basic}}}\big]
  \,=\,
  0
  \phantom{AAAA}
  \mbox{in situation \ref{The4SphericalFibration}}
  \,.
$$
This means, according to \eqref{TheAnomalyPolynomials}, that the total M5-brane anomaly is finally cancelled.

\end{enumerate}

\section{A resolution}
\label{TheSolution}

We now prove
that \hyperlink{HypothesisH}{\it Hypothesis H}
implies, in the situation \eqref{The4SphericalFibration},
the vanishing of the problematic basic term
$\big[G_4^{{}^{\mathrm{basic}}}\big]$ in \eqref{Density4Form},
thus implying the vanishing
of the total M5-brane anomaly according to \eqref{TheAnomalyPolynomials}.
We proceed in two steps:
\begin{enumerate}[{\bf (1)}]
\setlength\itemsep{-.08cm}
\item
In \cref{ViaGenericTadpoleCancellation}, we observe a general mechanism
that applies as soon as the Bianchi identity
$d G_7 = - \tfrac{1}{2} G_4 \wedge G_4$ holds with any correction by
Pontrjagin classes (which is traditionally not guaranteed, see Rem. \ref{OnFurtherCorrectionsToTheBianchiIdentity}).

\item
In \cref{ViaHypothesisH}, we discuss how \hyperlink{HypothesisH}{Hypothesis H}
implements this mechanism.
\end{enumerate}

\noindent
Both steps rely on facts about tangent structure on sphere bundles,
whose proofs we relegate to \cref{TangentStructuresOnSphereBundles}.

\subsection{For generic $G_7$-Bianchi identity}
\label{ViaGenericTadpoleCancellation}

\begin{theorem}[Vanishing of basic component]
  \label{BasicComponentVanishes}
  Given a black M5-brane background \eqref{The4SphericalFibration}
  with C-field flux $G_4$ \eqref{Density4Form}
  satisfying a Bianchi identity of the form
  \begin{equation}
    \label{GenericBianchiIdentity}
    d\, G_7
    \;=\;
    -
    \tfrac{1}{2}
    G_4 \wedge G_4
    \;\;+\;\;
    P\big(
      p_1(\nabla^{{}^{T X}}),
      p_2(\nabla^{{}^{T X}})
    \big)
    \;\;\;\;\;
    \in
    \Omega^8_{\mathrm{dR}}
    (
      X
    )
  \end{equation}
  for $P$ \emph{any} polynomial of
  Pontrjagin forms, then the basic component of $[G_4]$  \eqref{Density4Form}
  vanishes:
  \begin{equation}
    \label{VanishingOfBasic4Form}
    \big[
      G_4^{{}^{\mathrm{basic}}}
    \big]
    \;=\;
    0
    \;\;\in\;
    H^4
    (
      B;
      \mathbb{R}
    )
    \,.
  \end{equation}
\end{theorem}
\begin{proof}
The key point is that
all Pontrjagin forms on a manifold that is an
orthogonal spherical fibration are basic forms,
by Prop. \ref{StableCharacteristicClassesOnSphereBundlesAreBasic}.
This means with \eqref{GenericBianchiIdentity} that
also the cup-square of the class of the 4-flux is basic:
\begin{equation}
  \label{EquationForG4CupCube-b}
  [G_4]^2
  \;=\;
  S(p)^\ast
  \Big(
    P
    \big(
      p_1(\NormalBundle),
      p_2(\NormalBundle)
    \big)
  \Big)
  \;\in\;
  H^{8}( X; \mathbb{R})
  \,.
\end{equation}
Consider then the fiber integration
\begin{equation}
  \label{FiberIntegration}
  S(p)_\ast
  \;:\;
  H^\bullet( X; \mathbb{R})
  \longrightarrow
  H^{\bullet-4}(\BaseSpace; \mathbb{R})
\end{equation}
along the fibers of $S(p)$,
as in \eqref{The4SphericalFibration}.
By \cite[Lemma 2.1]{BottCattaneo97},
the fiber integration of the odd cup power $\rchi_4^{2k+1}$ of the
Euler class $\rchi_4 \in H^4( X; \mathbb{R})$ of the fibration $S(p)$ are proportional
to cup powers of the second Pontrjagin class of $\NormalBundle$:
\begin{equation}
  \label{FiberIntegrationOfPowersOfEulerClassesOverSpheres}
  S(p)_\ast
  (
    \rchi_4^{2k+1}
  )
  =
  2
  \big(
    p_2
    (
      \NormalBundle
    )
  \big)^k
  \;\;\in\;\;
  H^{8k}(\BaseSpace)
  \,,
\end{equation}
while the fiber integration of the even cup powers
of the Euler class vanishes for all $k \in \mathbb{N}$:
\begin{equation}
  \label{FiberIntegrationOfEvenPowersOfEulerClassesOverSpheres}
  S(p)_\ast
  (
    \rchi_4^{2k}
  )
  =
  0
  \;\in\;
  H^{ 8k-4 }(\BaseSpace)\;.
\end{equation}
Notice also the {\it projection formula} (e.g. \cite[Prop. 6.16]{BottTu82}\cite[(2)]{FSS18})
\begin{equation}
  \label{ProjectionFormula}
  S(p)_*
  \Big(
    \big(
      S(p)^*\alpha
    \big)
    \wedge
    \beta
  \Big)
    \;=\;
  \alpha
  \wedge
  S(p)_*\beta
  \,,
\end{equation}
which in particular implies that the fiber integral of basic forms vanishes:
\begin{equation}
  \label{FiberIntegrationOfBasicFormVanishes}
  S(p)_* S(p)^* \alpha
  \;=\;
  S(p)_*
  \Big(
    S(p)^*\alpha
    \wedge
    1
  \Big)
  \;=\;
  \alpha
  \wedge
  S(p)_* 1
  \;=\;
  0
  \,.
\end{equation}
Therefore, from
(\ref{EquationForG4CupCube-b})
and by repeated use of formulas
(\ref{FiberIntegrationOfPowersOfEulerClassesOverSpheres}
-
\ref{FiberIntegrationOfBasicFormVanishes})
we get:
\begin{equation}
\label{ComputationOfVanishingBasicComponent}
\begin{aligned}
  0
  & =
  S(p)_* S(p)^*
  P
  \big(
    p_1(\NormalBundle),
    p_2(\NormalBundle)
  \big)
  \\
  &
  =
  \tfrac{1}{2}
  S(p)_*
  [G_4 \wedge G_4]
  \\
  &
  =
  \tfrac{1}{2}
  S(p)_*
  \Big(
    \big(
      \tfrac{1}{2}
      \rchi_4
      +
      S(p)^*[G_4^{\mathrm{basic}}]
    \big)
    \wedge
    \big(
      \tfrac{1}{2}
      \chi
      +
      S(p)^*[G_4^{\mathrm{basic}}]
    \big)
  \Big)
  \\
  &=
  \tfrac{1}{2}
  S(p)_*(\rchi_4^2)
  +
  S(p)_*
  \big(
    \tfrac{1}{2}\rchi_4
    \wedge
    S(p)^*[G_4^{\mathrm{basic}}]
  \big)
  +
  \tfrac{1}{2}
  S(p)_*
  \big(
    S(p)^*[G_4^{\mathrm{basic}}]
  \big)^2
  \\
  &
  =
  \tfrac{1}{2}
  S(p)_*
  \big(
    \rchi_4^2
  \big)
  +
  S(p)_*
  \big(
    \tfrac{1}{2}\rchi_4
  \big)
   \wedge
  [G_4^{\mathrm{basic}}]
  +
  \tfrac{1}{2}
  S(p)_* S(p)^*
  [G_4^{\mathrm{basic}} \wedge G_4^{\mathrm{basic}}]
  \\
  &
  =
  [G_4^{\mathrm{basic}}]
  \,.
\end{aligned}
\end{equation}

\vspace{-5mm}
\hfill \end{proof}

\begin{remark}[M5-Brane anomaly cancellation for traditional $G_7$-Bianchi identity]
  \label{OnTheTraditionalBianchiIdentity}
A Bianchi identity for the (Hodge) dual flux density
$G_7 \coloneqq *_{{}_{11}} G_4$ of the form
\eqref{GenericBianchiIdentity} is traditionally considered as
(\cite[(1.1)]{DuffLiuMinasian95}\cite[(7.2)]{DFM03})
\begin{equation}
  \label{TraditionalBianchiIdentity}
  d \, G_7
  \;=\;
  - \tfrac{1}{2} G_4^2
  \;+\;
  I_8
  \big(
    p_1(\nabla^{{}^{T X}}),
    p_2(\nabla^{{}^{T X}})
  \big)
  \,
\end{equation}
for the polynomial $P$ being just the $I_8$-term in \eqref{TheAnomalyPolynomials}:
\begin{equation}
  \label{TheI8Term}
  I_8(p_1, p_2)
  \;\coloneqq\;
  \tfrac{1}{48}
  \big(
    p_2 - \tfrac{1}{4}p_1^2
  \big)
  \,.
\end{equation}
Under this traditional assumption,
Theorem \ref{BasicComponentVanishes}
implies the condition \eqref{TheAnsatz}
and hence the vanishing of the remaining M5-brane anomaly
in \eqref{TheAnomalyPolynomials}.
\end{remark}
\begin{remark}
  \label{ComparisonToBBMN19b}
  Theorem \ref{BasicComponentVanishes} in conjunction with
  Remark \ref{OnTheTraditionalBianchiIdentity}
  may be compared to an analogous
  physics argument in \cite[\S 4.1]{BahBonettiMinasianNardoni19b}
  (which appeared after \cite{FSS19a}
  and during the writing of the first version of this article;
  we thank an anonymous referee for pointing this out).

\end{remark}

\begin{remark}[Further corrections to the $G_7$-Bianchi identity]
\label{OnFurtherCorrectionsToTheBianchiIdentity}
The traditional Bianchi identity \eqref{TraditionalBianchiIdentity}
incorporates only one of several
expected corrections to the plain supergravity Bianchi identity ($P = 0$).
These M-theoretic {\it higher curvature corrections} are
traditionally investigated via an action principle
(e.g. \cite{HoweTsimpis03}\cite{HyakutakeOgushi06}\cite{SoueresTsimpis17}):

\vspace{-3mm}
\begin{enumerate}[{\bf (i)}]
\setlength\itemsep{-.08cm}
\item
From the action principle one
expects further higher derivative contributions to the Bianchi identity
\eqref{TraditionalBianchiIdentity} (\cite[(4.11)]{SoueresTsimpis17}, following \cite[around (56)]{HoweTsimpis03}):

\vspace{-.7cm}
\begin{equation}
  \label{HigherCurvatureCorrectedBianchiIdentity}
  d \, G_7
  \;=\;
  - \tfrac{1}{2} G_4^2
  \;+\;
  I_8
  (
    p_1,
    p_2
  )
  \;+\;
  \overset{
    \mathclap{
    \mbox{
      \tiny
      \color{darkblue}
      \bf
      \begin{tabular}{c}
        further
        \\
        corrections
      \end{tabular}
    }
    }
  }
  {
    \delta \Delta L
  }
  \,,
\end{equation}
  which locally, on a chart $U \subset X$ where $G_4\big\vert_{U} = d C_3^U$
  (\cite[below (3.4)]{SoueresTsimpis17}),
  are of the form (\cite[below (4.11)]{SoueresTsimpis17})
\begin{equation}
   \delta \Delta L\big\vert_U
   \;=\;
   d
   \big(
     \tfrac{\delta}{\delta C_3^U}
     \cdots
   \big)
   \,.
\end{equation}
This shows at once that:
\vspace{-3mm}
\begin{enumerate}[{\bf (a)}]
\setlength\itemsep{-.06cm}
\item {\it locally on $U$} the correction is exact
(which is the case highlighted in \cite[below (4.11)]{SoueresTsimpis17} ),
but
\item  {\it globally on $X$} it fails to be exact as soon as
$G_4$ is not globally exact (is only the curvature 4-form of
a 2-gerbe connection with local connection 3-forms $\{C_3^U\}_{U \in \mathcal{U}}$
on an open cover $\mathcal{U}$ of $X$, e.g. \cite[p. 22]{FSS12b}),
which is the generic case and the case of interest here, due to \eqref{UnitFluxCondition}.
\end{enumerate}

\item The action principle, and hence any Bianchi identity
derived from it, must moreover incorporate a global shift
\cite[(4.16)]{Tsimpis04}:
\begin{equation}
  \label{ShiftInActionPrinciple}
  S(g, G_4) \;\longmapsto\; S\big(g, G_4 + \tfrac{1}{4} p_1(\nabla^{{}^{T X}}) \big)
  \,,
\end{equation}
reflecting the
expected \cite[(1.2)]{Witten96a}[(1.2)]\cite{Witten96}
shifted flux quantization
\eqref{ShiftedIntegralityCondition} of the C-field.
\end{enumerate}
Apart form the general question of whether a classical action principle,
of all things,
can be the right principle for resolving foundations of M-theory,
the complete form of the higher curvature correction in
\eqref{HigherCurvatureCorrectedBianchiIdentity} remains open, and
its combination with the shift \eqref{ShiftInActionPrinciple}
in the action principle seems not to have been discussed yet.
Hence, under traditional assumptions, it remains unknown
whether the assumption \eqref{GenericBianchiIdentity}
is met in full M-theory.
\end{remark}
\medskip
What has been missing is a principle that fixes
Bianchi identities on more fundamental grounds.
Such a principle is {\it cohomological flux quantization}
\eqref{GeneralTwistedCharacterMap}, to which we turn now
in \cref{ViaHypothesisH}.

\medskip

\subsection{Via Hypothesis H}
\label{ViaHypothesisH}

We briefly recall from \cite{FSS19b}\cite{FSS20b} the
motivation and formulation of \hyperlink{HypothesisH}{\it Hypothesis H}
on the flux quantization principle for the C-field in M-theory.
Then we show (Thm. \ref{AnomalyCancellationByHypothesisH} below)
how this Hypothesis implements
the above mechanism for cancellation of the remaining M5-anomaly term.

\medskip

\noindent
{\bf Flux/charge quantization of higher gauge fields.}
The key idea is that the mathematical nature of any higher gauge field
is encoded in a {\it twisted generalized cohomology theory} $\widetilde A{}^\tau(-)$,
a notion known
as {\it flux quantization} or {\it charge quantization} (see \cite{Freed00}\cite{tcu}\cite{FSS20c}):
A generalized twisted {\it character map} \cite[\S 5]{FSS20c}
approximates cocycles in $\tau$-twisted
$A$-cohomology by flux densities in
twisted $L_\infty$-algebra valued de Rham cohomology,
namely by differential forms satisfying
polynomial differential relations --  {\it Bianchi identities}:
\begin{equation}
\label{GeneralTwistedCharacterMap}
\begin{tikzcd}[row sep=0pt]
  \overset{
   \mathclap{
    \raisebox{4pt}{
      \tiny
  \bf    \color{darkblue}
      \bf
      \begin{tabular}{c}
        $\tau$-twisted $A$-cohomology
        \\
        of spacetime $X$
      \end{tabular}
    }
    }
  }{
    \widetilde A{}^\tau
    (X)
  }
  \ar[
    rr,
    "
      \overset{
        \raisebox{4pt}{
          \tiny
          \color{greenii}
          \bf
          \begin{tabular}{c}
            twisted character map
          \end{tabular}
        }
      }{
        \mathrm{ch}^\tau_A
      }
    "
  ]
  &{\phantom{AAA}}&
  \overset{
    \mathclap{
    \raisebox{4pt}{
      \tiny
      \color{darkblue}
      \bf
      \begin{tabular}{c}
        $\tau_{\mathrm{dR}}$-twisted
        \\
        $L_\infty$-valued de Rham cohomology
      \end{tabular}
    }
    }
  }{
    H^{\tau_{\mathrm{dR}}}_{\mathrm{dR}}
    \big(
      X;
      \mathfrak{l}A
    \big)
  }
  \\
  \overset{
    \mathclap{
    \raisebox{3pt}{
      \tiny
      \color{darkblue}
      \bf
      \begin{tabular}{c}
        $\tau$-twisted $A$-cocycle
      \end{tabular}
    }
    }
  }{
    \big[
      X
      \xrightarrow{c}
     A \!\sslash\! G
    \big]
  }
  \ar[
    rr,
    phantom,
    shift left=3pt,
    "
      \longmapsto
    "{
      description
    },
    "
        \mathclap{
        \raisebox{4pt}{
          \tiny
          \color{greenii}
          \bf
          rational approximation
        }
        }
    "{
      above
    }
  ]
  \ar[
    rr,
    phantom,
    shift right=3pt,
    "
      \rotatebox{-180}{
        \scalebox{1}{$
          \longmapsto
        $}
      }
    "{
      description
    },
    "
        \mathclap{
        \raisebox{-6pt}{
          \tiny
          \color{greenii}
          \bf
          flux quantization
        }
        }
    "{
      below
    }
  ]
  &&
  \overset{
    \hspace{-45pt}
    \mathclap{
    \raisebox{4pt}{
      \tiny
      \color{darkblue}
      \bf
      \begin{tabular}{c}
        flux densities
      \end{tabular}
    }
    }
  }{
  \big(
    F_i(c) \in
    \Omega^\bullet_{\mathrm{dR}}(X)
  \big)_{i \in I}
  }
  \Big\vert
  \Big(
    d\, F_i(c)
    \overset{
      \mathclap{
      \raisebox{8pt}{
        \tiny
        \color{darkblue}
        \bf
        \begin{tabular}{c}
          Bianchi identities
        \end{tabular}
      }
      }
    }{
      =
    }
    P_i
    \big(
      \{F_j(c)\}_{j \in I},
      \overset{
        \mathclap{
        \raisebox{+4pt}{
          \tiny
          \color{darkblue}
          \bf
          \begin{tabular}{c}
            background
            \\
            fluxes
          \end{tabular}
        }
        }
      }{
        \;\;
        \tau_{\mathrm{dR}}
        \;\;
      }
    \big)
  \Big)_{i \in I}
\end{tikzcd}
\end{equation}
Flux/charge quantization in $A$-theory means to demand that the
flux densities are in the image of the twisted $A$-character
map \eqref{GeneralTwistedCharacterMap}
of an actual cocycle $c$ in twisted $A$-cohomology,
which then embodies the actual field configuration
(its topological sector as shown here, for brevity, and the
full field configuration after refinement to {\it differential} $A$-cohomology
\cite[\S 4.3]{FSS20c}).

\medskip
The archetypical example is Dirac's flux quantization of the electromagnetic field
(e.g. \cite[\S 2]{Alvarez85}\cite[\S 2]{Freed00}),
which is the demand that the ordinary electromagnetic flux density $F_2$ (the Faraday tensor)
is the character image of a cocycle in ordinary integral degree-2 cohomology
$A(-) = H^2(-;\mathbb{Z})$ (hence is the curvature 2-form of a
connection on a complex line bundle), which equivalently means that
it represents an integral cohomology class
$[F_2] \in H^2(X;\mathbb{Z}) \xrightarrow{\;} H^2(X;\mathbb{R})$.
Here the Bianchi identity
obtained from \eqref{GeneralTwistedCharacterMap}
is the simple closure condition $d F_2 \,=\, 0$
(\cite[Ex. 4.10]{FSS20c}).

\medskip
The most famous example is the K-theory conjecture
in string theory
\cite{MinasianMoore97}\cite{Witten98}\cite{Freed00}
which states
(review in \cite{Witten00}\cite{Fredenhagen08})
that the B-field and the RR-field fluxes in type II string theory are jointly
quantized in twisted \cite[\S 5.3]{Witten98}\cite{BouwknegtMathai00}
(and differential, see \cite{GS-RR} for recent rigorous developments)
topological K-theory, $A^\tau(-) = \mathrm{KU}^\tau(-)$.
Indeed, the character map \eqref{GeneralTwistedCharacterMap}
in this case takes the following form
(\cite[\S 2.5]{FreedHopkinsTeleman02}\cite[Prop. 5.5]{FSS20c},
shown here for type IIA string theory, for definiteness):
\begin{equation}
  \label{TwistedChernCharacter}
  \underset{
    \mathclap{
    \raisebox{-4pt}{
      \tiny
      \color{darkblue}
      \bf
      \begin{tabular}{c}
        $\tau$-twisted
        \\
        complex K-theory
        \\
        of $X$
      \end{tabular}
    }
    }
  }{
    \mathrm{KU}^\tau(X)
  }
  \;\coloneqq\;
  \left\{
  \begin{tikzcd}
    &
    \mathllap{
    \mathbb{Z} \times
    }
    B\mathrm{U} \!\sslash\! B \mathrm{U}(1)
    \ar[
      d
    ]
    \\
    X
    \ar[
      r,
      "\tau"{
        above
      },
      "
        \mbox{
          \tiny
          \color{greenii}
          \bf
          \begin{tabular}{c}
            classifying map
            \\
            of B-field
          \end{tabular}
        }
      "{
        below
      }
    ]
    \ar[
      ur,
      dashed,
      "
        \overset{
          \mathllap{
          \raisebox{3pt}{
            \tiny
            \color{greenii}
            \bf
            \begin{tabular}{c}
              classifying map
              \\
              of RR-fields
            \end{tabular}
          }
          }
        }{
          r
        }
        \;\;\;\;\;\;\;\;
      "{
        left
      },
      near end
    ]
    &
    B^2 \mathrm{U}(1)
  \end{tikzcd}
    \!\!\!\!\!
  \right\}_{
    \!\!\!\!\!
    \big/
    {
      \mathrm{vertical}
        \atop
      \mathrm{homotopy}
    }
  }
    \hspace{-40pt}
  \begin{tikzcd}
    {\phantom{A}}
    \ar[
      rr,
      "
        \overset{
          \mathclap{
          \raisebox{3pt}{
            \tiny
            \color{greenii}
            \bf
            \begin{tabular}{c}
              twisted
              \\
              Chern character
            \end{tabular}
          }
        }
      }{
        \mathrm{ch}^\tau_{\mathrm{KU}}
      }
      ",
      "
        \mbox{
         \tiny
        }
      "{
       below
      }
    ]
    &&
    {\mathclap{\phantom{A}}}
  \end{tikzcd}
  \!\!\!\!
    \left\{
      \!\!\!\!
      \left.
      \def\arraystretch{1.3}
      \begin{array}{c}
        H_3\mathrlap{,}
        \\
        \{F_{2k}\}_k
      \end{array}
      \!
      \right\vert
      \!\!
      \raisebox{0pt}{$
      \def\arraystretch{1.3}
      {\begin{array}{ll}
        d\, H_3
          &
          \!\! = 0
          \,,
          \;\;\;
          \big[
            \overset{
              \mathclap{
              \raisebox{3pt}{
                \tiny
                \color{darkblue}
                \bf
                \begin{tabular}{c}
                  NS 3-flux
                \end{tabular}
              }
              }
            }{
            H_3
            }
          \big]
          \in H^3(X;\mathbb{Z}) 
        \\
        \underset{
          \mathrlap{
          \;\;\;\;\;\;\;\;\;\;\;\;\;\;\;
          \raisebox{-3pt}{
            \tiny
            \color{darkblue}
            \bf
            RR-flux Bianchi identities
          }
          }
        }{
          d\, F_{2k+2}
        }
          &
          \!\!=
          H_3 \wedge F_{2k}
      \end{array}}
    $}
    \!\!\!
    \right\}_{
     \mathrlap{
      \!\!\!
      \big/\sim
      }
    }
\end{equation}
The Bianchi identities
on the right of \eqref{TwistedChernCharacter}
are exactly those expected to be satisfied by the NS B-field flux $H_3$
and the RR-flux densities $F_{2k}$ in type IIA string theory
(see \cite[\S 4]{FSS16b}\cite[\S 1]{BMSS19} for details and pointers).

\medskip
But the M-theoretic lift of these $H_3/F_{2k}$-Bianchi identities \eqref{TwistedChernCharacter}
is (see \cite[\S 4.2]{MathaiSati03}\cite{tcu}\cite[\S 3]{FSS16a}\cite[\S 4]{BMSS19})
just the $G_7$-Bianchi identity \eqref{GenericBianchiIdentity}
together with closure and the shifted integrality condition on the $G_4$-flux:

\vspace{-5mm}
\begin{equation}
  \label{ShiftedIntegralityCondition}
  d\, G_4 = 0
  \,,
  \;\;\;\;
  \big[
    \overset{
     \widetilde G_4
    }{
    \overbrace{
      G_4 + \tfrac{1}{4}p_1(\nabla^{T X})
    }
    }
  \big]
  \;
  \in
  \;
  H^4(X; \mathbb{Z}) \xrightarrow{\;} H^4(X; \mathbb{R})
  \,.
\end{equation}
Therefore, it is  natural to ask for a cohomology theory
whose twisted character map \eqref{GeneralTwistedCharacterMap} enforces \eqref{GenericBianchiIdentity}
and \eqref{ShiftedIntegralityCondition}:

\medskip

\newpage

\hypertarget{HypothesisH}{}
\noindent
{\bf Hypothesis H} is the statement
\cite[\S 2]{FSS19b}\cite{FSS19c}\cite{SS19a}\cite{SS21}
(following \cite[\S 2.5]{Sati13}\cite{FSS16a}, review in \cite[\S 7]{FSS19})
that  the cohomology theory for flux/charge quantization \eqref{GeneralTwistedCharacterMap}
of the
C-field in M-theory is Borsuk-Spanier {\it Cohomotopy theory}
$A^\tau(-) = \pi^\tau(-)$ \cite{Spanier49}
in joint degrees 4 (for the M5-brane charge) and 7 (for the M2-brane charge)
related by the quaternionic Hopf fibration  and
twisted by the tangent bundle
via the J-homomorphism (``J-twist''):

\medskip
The classifying space for degree-$n$ Cohomotopy is (the homotopy type of)
the $n$-sphere $S^n$, and for (orthogonally) twisted $n$-Cohomotopy it is
the homotopy quotient (Borel construction) $S^n \sslash \mathrm{O}(n+1)$
of the canonical action of
the orthogonal group $\mathrm{O}(n+1)$ on $S^n \simeq S(\mathbb{R}^{n+1})$
(recalled as Def. \ref{OrthonalSphereFiberBundle} below):
\begin{equation}
  \label{TwistedCohomotopicalCharacter}
  \underset{
    \mathclap{
    \raisebox{-4pt}{
      \tiny
      \color{darkblue}
      \bf
      \begin{tabular}{c}
        $\tau$-twisted
        \\
        4-Cohomotopy
        \\
        of $X$
      \end{tabular}
    }
    }
  }{
    \pi^\tau(X)
  }
  \;\coloneqq\;
  \left\{
  \!\!\!
  \begin{tikzcd}
    &
    S^4 \!\sslash\! \mathrm{O}(5)
    \ar[
      d
    ]
    \\
    X
    \ar[
      r,
      "\tau"{
        below
      }
    ]
    \ar[
      ur,
      dashed,
      "
        \overset{
          \mathllap{
          \raisebox{3pt}{
            \tiny
            \color{greenii}
            \bf
            cocycle
          }
          }
        }{
          c
        }
        \;\;\;\;\;\;\;\;
      "{
        left
      },
      near end
    ]
    &
    B \mathrm{O}(5)
  \end{tikzcd}
  \!\!\!\!\!
  \right\}_{
    \!\!\!\!\!
    \big/
    {
      \mathrm{vertical}
        \atop
      \mathrm{homotopy}
    }
  }
  \hspace{-40pt}
  \begin{tikzcd}
    {\phantom{A}}
    \ar[
      rr,
      "
        \overset{
          \mathclap{
          \raisebox{3pt}{
            \tiny
            \color{greenii}
            \bf
            \begin{tabular}{c}
              twisted
              \\
              cohomotopical
              \\
              character
            \end{tabular}
          }
        }
      }{
        \mathrm{ch}^\tau_\pi
      }
      ",
      "
        \mbox{
         \tiny
         \cite[\S 5.3]{FSS20c}
        }
      "{
       below
      }
    ]
    &&
    {\mathclap{\phantom{A}}}
  \end{tikzcd}
  \!\!\!\!
    \left\{
      \!\!\!\!
      \left.
      \def\arraystretch{1.3}
      \begin{array}{c}
        G_4\mathrlap{,}
        \\
        G_7
      \end{array}
      \!
      \right\vert
      \!\!
      \def\arraystretch{1.3}
      {\begin{array}{ll}
        d\, G_4
          &
          \!\! = 0
          \,, \;
          \big[
            G_4 + \tfrac{1}{4}p_1(\nabla^\tau)
          \big]
          \in H^4(X;\mathbb{Z}) 
        \\
        d\, G_7
          &
          \!\!=
          - \tfrac{1}{2} G_4 \wedge G_4 + \tfrac{1}{8} p_2(\nabla^\tau)
      \end{array}}
    \!\!\!
    \right\}_{
      \!\!\!
      \big/\sim
    }
\end{equation}
On the right of \eqref{TwistedCohomotopicalCharacter} we are showing the
form of the image of the character map \eqref{GeneralTwistedCharacterMap}
specified to orthogonally twisted 4-Cohomotopy
(due to \cite[Prop. 2.5 \& 3.13]{FSS19b}, for more see \cite[\S 5.3]{FSS20c}).
Both the $\tfrac{1}{4} p_1$-shifted integral flux quantization on $G_4$
\eqref{ShiftedIntegralityCondition}
is implied from charge-quantization in twisted Cohomotopy,
as well as
the general form of the $G_7$-Bianchi identity \eqref{GenericBianchiIdentity}.
It just remains to relate the Pontrjagin classes of the twisting bundle
$\tau$ to the tangent bundle:

\medskip
The condition that the twist $\tau$ be compatible in degrees 7 and 4, along
the quaternionic Hopf fibration $h_{\mathbb{H}}$ singles out
(\cite[Prop. 2.20]{FSS19b})
the  quaternionic central product subgroup
$\mathrm{Sp}(2) \cdot \mathrm{Sp}(1) \subset \mathrm{Spin}(8) \to \mathrm{O}(8)$;
and demanding that it, moreover, be compatible with factorization through the
Atiyah-Penrose twistor fibration $t_{\mathbb{H}}$
(which corresponds \cite{FSS20b}\cite{SS20c} to charge-quantization in
{\it heterotic} M-theory)
singles out
(\cite[Prop. 2.2]{FSS20b})
the further subgroup
$\mathrm{Sp}(2) \,\subset\, \mathrm{Sp}(2) \cdot \mathrm{Sp}(1)$:
\begin{equation}
  \begin{tikzcd}
    S^7
  \ar[
    out=180-66,
    in=66,
    looseness=3.5,
    "
    \scalebox{.77}{$
      \phantom{\cdot}
      \mathclap{
        \mathrm{Sp}(2)\cdot\mathrm{Sp}(1)
      }
      \phantom{\cdot}
    $}
    "{
      description
    },
    shift right=1
  ]
    \ar[
      rr,
      "h_{\mathbb{H}}"{
        above
      },
      "
        \mbox{
          \tiny
          \color{greenii}
          \bf
          \begin{tabular}{c}
            quaternionic
            \\
            Hopf fibration
          \end{tabular}
        }
      "{
        below
      }
    ]
    &&
    S^4
    \mathrlap{
      \,,
    }
  \ar[
    out=180-66,
    in=66,
    looseness=3.5,
    "
    \scalebox{.77}{$
      \phantom{\cdot}
      \mathclap{
        \mathrm{Sp}(2)\cdot\mathrm{Sp}(1)
      }
      \phantom{\cdot}
    $}
    "{
      description
    },
    shift right=1
  ]
  \end{tikzcd}
  {\phantom{AAAAAAA}}
  \begin{tikzcd}
    S^7
  \ar[
    out=180-66,
    in=66,
    looseness=3.5,
    "
    \scalebox{.77}{$
      \phantom{\cdot}
      \mathclap{
        \mathrm{Sp}(2)
      }
      \phantom{\cdot}
    $}
    "{
      description
    },
    shift right=1
  ]
    \ar[
      rr,
      "h_{\mathbb{C}}"{
        above
      },
      "
        \mbox{
          \tiny
          \color{greenii}
          \bf
          \begin{tabular}{c}
            complex
            \\
            Hopf fibration
          \end{tabular}
        }
      "{
        below
      }
    ]
    &&
    \mathbb{C}P^3
  \ar[
    out=180-66,
    in=66,
    looseness=3.5,
    "
    \scalebox{.77}{$
      \phantom{\cdot}
      \mathclap{
        \mathrm{Sp}(2)
      }
      \phantom{\cdot}
    $}
    "{
      description
    },
    shift right=1
  ]
    \ar[
      rr,
      "t_{\mathbb{H}}"{
        above
      },
      "
        \mbox{
          \tiny
          \color{greenii}
          \bf
          \begin{tabular}{c}
            Atiyah-Penrose
            \\
            twistor fibration
          \end{tabular}
        }
      "{
        below
      }
    ]
    &&
    S^4
    \,.
  \ar[
    out=180-66,
    in=66,
    looseness=3.5,
    "
    \scalebox{.77}{$
      \phantom{\cdot}
      \mathclap{
        \mathrm{Sp}(2)
      }
      \phantom{\cdot}
    $}
    "{
      description
    },
    shift right=1
  ]
  \end{tikzcd}
\end{equation}

A key subtlety here is that the quaternionic unitary group
$\mathrm{Sp}(2)$ and the spin-group $\mathrm{Spin}(5)$
are isomorphic as abstract Lie groups, but not as subgroups of
$\mathrm{Spin}(8) \to \mathrm{O}(8)$ (nor are they conjugate subgroups);
instead (\cite[Prop. 2.17]{FSS19b}) as subgroups they are mapped to each other
under the triality automorphism on the ambient $\mathrm{Spin}(8)$-group:
\begin{equation}
 \label{TrialityAutomorphism}
  \begin{tikzcd}[row sep=0, column sep=35pt]
    \mathrm{Sp}(2)
    \ar[
      dd,
      hook,
      "i_{\mathrm{Sp}(2)}"{
        left
      }
    ]
    \ar[
      rr,
      "\sim"
    ]
    &&
    \mathrm{Spin}(5)
    \ar[
      dd,
      hook,
      "i_{\mathrm{Spin}(5)}"
    ]
    \\
    {\phantom{A}}
    \\
    \mathrm{Spin}(8)
    \ar[
      rr,
      "\sim"{
       below
      },
      "
       \overset{
         \mathclap{
         \raisebox{+3pt}{
           \tiny
           \color{greenii}
           \bf
           triality automorphism
         }
         }
       }{
          \mathrm{tri}
       }
      "{
        above
      }
    ]
    &&
    \mathrm{Spin}(8)
    \mathrlap{
      \,,
    }
    \\
    {\phantom{
      \scalebox{.7}{$
        p_1
      $}
    }}
    \\
    {\phantom{
      \scalebox{.7}{$
        -
        \big(
          \tfrac{1}{4}
          p_1
        \big)^2
        +
        12 I_8
      $}
    }}
  \end{tikzcd}
  {\phantom{AAAAA}}
  \begin{tikzcd}[row sep=0pt, column sep=40pt]
    B \mathrm{Sp}(2)
    \ar[
      dd,
      hook,
      "B i_{\mathrm{Sp}(2)}"{
        left
      }
    ]
    \ar[
      rr,
      "\sim"
    ]
    &&
    B \mathrm{Spin}(5)
    \ar[
      dd,
      hook,
      "B i_{\mathrm{Spin}(5)}"
    ]
    \\
    {\phantom{A}}
    \\
    B \mathrm{Spin}(8)
    \ar[
      rr,
      "\sim"{
       below
      },
      "
       \overset{
         \mathclap{
         \raisebox{+3pt}{
           \tiny
           \color{greenii}
           \bf
           delooped triality automorphism
         }
         }
       }{
          B \mathrm{tri}
       }
      "{
        above
      }
    ]
    &&
    B \mathrm{Spin}(8)
    \\
    \scalebox{.7}{$
      p_1
    $}
    &
    \scalebox{.7}{$
      \longleftrightarrow
    $}
    &
    \scalebox{.7}{$
      p_1
    $}
    \\
    \scalebox{.7}{$
      \big(
        \tfrac{1}{4}p_1
      \big)^2
      -
      24 \cdot I_8
    $}
    &
    \scalebox{.7}{$
      \longleftrightarrow
    $}
    &
    \scalebox{.7}{$
      \tfrac{1}{4} p_2
    $}
  \end{tikzcd}
\end{equation}
As shown on the bottom right of \eqref{TrialityAutomorphism},
this triality automorphism,
does not affect the first Pontrjagin class,
but does induce a nontrivial transformation of
Pontrjagin classes in degree 8 (\cite[Lem. 2.19]{FSS19b}).
Therefore, as we consider
tangential $\mathrm{Sp}(2)\cdot \mathrm{Sp}(1)$-structure on spacetime
to unify M2/M5-brane charge quantization in J-twisted Cohomotopy,
and in fact tangential $\mathrm{Sp}(2)$-structure to
account for charges in heterotic-theory,
we arrive at J-twisted Cohomotopy theory
of the following form \cite[(17)]{FSS19b}\cite[(43)]{FSS19c}:
  \begin{equation}
  \label{JTwistedCohomotopy}
  \hspace{-2mm}
  \overset{
    \mathclap{
    \raisebox{5pt}{
      \tiny
      \color{darkblue}
      \bf
      \begin{tabular}{c}
        J-twisted
        \\
        4-Cohomotopy
      \end{tabular}
    }
    }
  }{
    \pi^{\tau_{\mathrm{Sp}(2)}}
    (
      X
    )
  }
      \coloneqq
    \left\{
  \;\;\;\;
  \begin{tikzcd}[column sep=4pt]
    &
    &
    \overset{
     \mathclap{
      \raisebox{7pt}{
        \tiny
        \color{darkblue}
        \bf
        \begin{tabular}{c}
          4-sphere bundle
          \\
          associated to
          \\
          $\mathrm{Sp}(2)$-reduced
          \\
          tangent bundle
        \end{tabular}
      }
      }
    }{
      E^4
    }
    \ar[
      dd
    ]
    \ar[
      rr
    ]
    \ar[
      ddrr,
      phantom,
      "
        \mbox{
         \tiny
         (pb)
        }
      "{
        description
      }
    ]
    &&
    \overset{
      \mathclap{
      \raisebox{7pt}{
        \tiny
        \color{darkblue}
        \bf
        \begin{tabular}{c}
          universal
          $\mathrm{Sp}(2)$-structured
          \\
          4-sphere bundle
        \end{tabular}
      }
      }
    }{
      S^4 \!\sslash\! \mathrm{Sp}(2)
    }
    \ar[
      rr,
      "\sim"
    ]
    \ar[dd]
    \ar[
      ddrr,
      phantom,
      "
        \mbox{
          \tiny
          \rm
          (pb)
        }
      "{
       description
      }
    ]
    &&
    S^4 \!\sslash\! \mathrm{Spin}(5)
    \ar[dd]
    \ar[rr]
    \ar[
      ddrr,
      phantom,
      "\mbox{\tiny\rm(pb)}"{
        description
      }
    ]
    &
    {\phantom{A}}
    &
    \overset{
      \raisebox{9pt}{
        \tiny
        \color{darkblue}
        \bf
        \begin{tabular}{c}
          universal
          \\
          orthogonal
          \\
          4-sphere bundle
        \end{tabular}
      }
    }{
      S^4 \!\sslash\! \mathrm{O}(5)
    }
    \ar[dd]
    \\
    \\
    X
    \ar[
      rr,
      -,
      shift left=1pt
    ]
    \ar[
      rr,
      -,
      shift right=1pt
    ]
    \ar[
      uurr,
      dashed,
      bend left=20,
      near end,
      "
        \overset{
         \mathllap{
          \raisebox{3pt}{
             \tiny
             \color{purple}
             \bf
             \begin{tabular}{c}
               cocycle in
               \\
               J-twisted
               \\
               Cohomotopy
             \end{tabular}
          }
          \;\;\;\;\;\;\;\;\;\;
          }
        }{
          c
          \;\;\;\;\;\;\;
        }
      "{
        left
      }
    ]
    &&
    X
    \ar[
      rr,
      "
        \mbox{
          \tiny
          \color{greenii}
          \bf
          \begin{tabular}{c}
            tangential
            $\mathrm{Sp}(2)$-structure
          \end{tabular}
        }
      "{
        above
      },
      "
        \tau_{\mathrm{Sp}(2)}
      "{
        below
      }
    ]
    \ar[
      drr,
      bend right=10,
      "
        \;\;\;\;\;\;
        \vdash T X
      "{
        above,
        near end
      },
      "
        \raisebox{-8pt}{
          \tiny
          \color{greenii}
          \bf
          \begin{tabular}{c}
            classifying map
            \\
            of tangent bundle
          \end{tabular}
        }
        \;\;\;\;\;\;\;\;\;\;\;\;\;\;\;
      "{
        below
      }
    ]
    &&
    B
    \mathrm{Sp}(2)
    \ar[
      rr,
      "\sim"
    ]
    \ar[d]
    &&
    B
    \mathrm{Spin}(5)
    \ar[d]
    \ar[rr]
    &&
    B \mathrm{O}(5)
    \\
    &{\phantom{A}}&
    &{\phantom{AAAAAA}}&
    B \mathrm{Spin}(8)
    \ar[
      rr,
      "
        B\mathrm{tri}
      "{
        below
      },
      "
       \mbox{
         \tiny
         \color{greenii}
         \bf
         \begin{tabular}{c}
           triality
           automorphism
         \end{tabular}
       }
      "{
        above
      }
    ]
    &
    {\phantom{AAAAA}}
    &
    B \mathrm{Spin}(8)
  \end{tikzcd}
\!\!\!\!\!\!\!\!\!  \right\}_{
    \mathrlap{
      \!\!\!\big/ { \mathrm{vertical} \atop \mathrm{homotopy}  }
    }
  }
\end{equation}
Under the twisted character map \eqref{GeneralTwistedCharacterMap}
(with
\eqref{TwistedCohomotopicalCharacter} and
\eqref{TrialityAutomorphism}), this implies
the following $G_7$-Bianchi identity \cite[Prop. 3.8]{FSS19b}\cite[\S 5.3]{FSS20c}:
\begin{equation}
  \label{G7BianchiIdentityFromHypothesisH}
\begin{aligned}
  d \,G_7
  & =
  -
  \tfrac{1}{2}
  \widetilde G_4
  \wedge
  \big(
    \widetilde G_4 - \tfrac{1}{2} p_1(\nabla^{{}^{T X}})
  \big)
  -
  12
  \cdot
  I_8
  \big(
    \nabla^{T X})
  \big)
  \,,
\end{aligned}
\end{equation}
for $\widetilde G_4 \coloneqq G_4 + \tfrac{1}{4}p_1$ \eqref{ShiftedIntegralityCondition}.

\begin{remark}[Structure of the Bianchi identity]
The Bianchi identity \eqref{G7BianchiIdentityFromHypothesisH}
is of the form \eqref{TraditionalBianchiIdentity} except for
inclusion of the integrality shift \eqref{ShiftedIntegralityCondition}
and of a relative weight on the
$I_8$-polynomial, corrections that are compatible with the
general expectations (Remark \ref{OnFurtherCorrectionsToTheBianchiIdentity}).
Detailed discussion of the consistency/necessity of these particular
corrections
for all of

{\bf (a)} C-field tadpole cancellation,

{\bf (b)} M5 WZ-term level-quantization,

{\bf (c)} M2-brane Page-charge quantization

\noindent
is given in \cite[p. 13 \& \S 3.8]{FSS19b}\cite{FSS19c}\cite[Rem. 4.1]{SS21}.
However, for the application to M5-brane anomaly cancellation,
these details are irrelevant. What matters here,
by Theorem \ref{BasicComponentVanishes}, is that the right hand side of
\eqref{G7BianchiIdentityFromHypothesisH} is proportional to
$G_4 \wedge G_4$ plus any polynomial in Pontrjagin forms.
\end{remark}

\noindent
{\bf M5-Brane anomaly cancellation via Hypothesis H.} This allows us to conclude:

\begin{theorem}
  \label{AnomalyCancellationByHypothesisH}
  If the base space $\BaseSpace$ is parallelizable
  and the normal bundle $\NormalBundle_{\mathrm{M5}}$ has
  $\mathrm{Sp}(2)$-structure
  then:
  \vspace{-2mm}
\begin{enumerate}[{\bf (i)}]
\setlength\itemsep{-.08cm}
\item the ambient black M5-brane spacetime \eqref{The4SphericalFibration}
  $X \xrightarrow{\;} \BaseSpace$
  carries
  tangential $\mathrm{Sp}(2)$-structure (Def. \ref{GStructure})
  $\tau_{\mathrm{Sp}(2)}$;

  \item
  flux-quantization \eqref{GeneralTwistedCharacterMap}
  of the C-field in
  $\tau$-twisted 4-Cohomotopy \eqref{JTwistedCohomotopy}
  enforces
  --
    besides the shifted 4-flux quantization \eqref{ShiftedIntegralityCondition}
    and
    the $G_7$-Bianchi identity \eqref{G7BianchiIdentityFromHypothesisH}
  --
  the vanishing of the class of $[G_4^{\mathrm{basic}}]$ \eqref{TheAnsatz}
  and hence of the remaining M5-brane anomaly \eqref{TheAnomalyPolynomials}.
\end{enumerate}
\end{theorem}
\begin{proof}
  By the exceptional coset space realization
  $
    S^4
      \,\simeq\,
      \mathrm{Sp}(2)/\big( \mathrm{Sp}(1) \times \mathrm{Sp}(1) \big)
  $
  \eqref{CosetSpaceRealizationsOfSpheres},
  Prop. \ref{HStructureOnVerticalTangentBundleOfGAssociatedSphereBundle}
  says that the vertical tangent bundle
  has
  $H$-structure, in particular $G$-structure, for $H \subset G$ being
  $\mathrm{Sp}(1) \times \mathrm{Sp}(1) \subset \mathrm{Sp}(2) \subset
  \mathrm{O}(8)$.
  By Prop. \ref{OnceStabilizedTangentBundleOfSphereBundleIsPulledBackFromBase}
  and using the assumption that the
  tangent bundle of $Q$ is trivializable,
  this is also the structure on the once-stabilized total tangent bundle,
  which is claim {(i)}.
  With this, claim (ii) follows with \eqref{G7BianchiIdentityFromHypothesisH}
  by Theorem \ref{BasicComponentVanishes}.
\hfill \end{proof}

\begin{remark}
  The assumption in Theorem \ref{AnomalyCancellationByHypothesisH}
  are met in the key examples of interest
  (see \cite{SS19a} \cite{FSS19d}\cite{FSS20a}
  further discussion and pointers):

  \noindent

  \vspace{-2mm}
  \begin{enumerate}[{\bf (i)}]
\setlength\itemsep{-.08cm}
  \item
  The assumption that the base space is parallelizable is satisfied for
  5-branes wrapped on tori
  $Q_{\mathrm{M5}} = \mathbb{R}^{5-n,1} \times T^n$
  or wrapped on the 3-sphere $Q_{\mathrm{M5}} = \mathbb{R}^{2,1} \times S^3$.

  \item
  The assumption that the normal bundle has $\mathrm{Sp}(2)$-structure
  is satisfied for 5-branes at ADE-singularities,
  where it even has $\mathrm{Sp}(1) \subset \mathrm{Sp}(2)$-structure.
  \end{enumerate}
\end{remark}

\newpage
\appendix

\section{Appendix: Tangent structure on sphere bundles}
\label{TangentStructuresOnSphereBundles}

Here we prove some results on
(vertical) tangent structures to sphere bundles.
The key consequences for the proofs in \cref{TheSolution} are:

\noindent {\bf (i)}
Prop. \ref{StableCharacteristicClassesOnSphereBundlesAreBasic}
(used in the proof of Thm.\ref{BasicComponentVanishes}
, $\;\;$

\noindent {\bf (ii)} Prop. \ref{HStructureOnVerticalTangentBundleOfGAssociatedSphereBundle}
(used in the proof of Thm. \ref{AnomalyCancellationByHypothesisH}).

\noindent The first of these must be well-known to experts, but complete
statements/proofs are hard to find in the literature
(we give commented pointers to existing references as we proceed).
We observe here that (i) follows as a direct corollary of
the second statement (ii), which seems to be new.
We give a slick homotopy-theoretic proof that neatly ties in
with the formulation of \hyperlink{HypothesisH}{\it Hypothesis H}.

\medskip

\noindent
{\bf Homotopy theory.}
Following \cite{FSS19b}\cite{FSS19c}\cite{SS21}, we make free use of basic notions of
homotopy theory (``higher structures'').
For mathematical background and pointers see \cite[\S A]{FSS20c}\cite[\S 2]{SS20b});
for exposition in the context of string/M-theory see
\cite{JSSW19}\cite{FSS19}.
This means that all topological spaces in the
following are regarded up to weak homotopy equivalence (see \cite[Ex. A.7]{FSS20c}),
which
we denote by an equality sign, e.g. for
$S^4 \!\sslash\! \mathrm{Spin}(5) = B \mathrm{Spin}(4)$ in \eqref{UniversalGAssociatedSphereBundle}
below,
where the double slash denotes the homotopy quotient
or {\it Borel construction} for any topological/simplicial group $G$
(see \cite[Prop. 3.73]{NSS12b})
$$
  X \!\sslash\! G
  \;=\;
  H \times_G E G
  \,,
$$
which subsumes the the classifying space construction $B(-)$
(e.g. \cite[Ex. 3.68]{NSS12b})
$$
  B G \;=\; \ast \!\sslash\! G
  \;=\;
  \ast \times_G E G
  \;=\;
  (E G)/G
$$
for {\it principal $G$-bundles}, being homotopy pullbacks
(e.g. \cite[A.24, A.27]{FSS20c}) of the universal $G$-principal bundle
$$
  E G
  \;\coloneqq\;
  G \!\sslash\! G
    \xrightarrow{\;}
  B G
$$
(\cite[\S 4.1]{NSS12b}\cite[\S 2.2]{SS20b}, traditional review in \cite{Mitchell11}):
\begin{equation}
  \label{ClassificationOfPrincipalBundles}
  \begin{tikzcd}[column sep=small]
    P
    \ar[
      rr
    ]
    \ar[
      d,
      "p"{
        right
      },
      "
        \mbox{
          \tiny
          \color{darkblue}
          \bf
          \begin{tabular}{c}
            principal
            \\
            $G$-bundle
          \end{tabular}
        }
      "{
        left
      }
    ]
    \ar[
      drr,
      phantom,
      "
        \overset{
          \raisebox{1pt}{
            \tiny
            \color{orangeii}
            \bf
            homotopy pullback
          }
        }{
        \mbox{
          \tiny\rm(pb)
        }
        }
      "
    ]
    &
    {\phantom{AAAA}}
    &
    G \!\sslash\! G
    \ar[r,-,shift left=1pt]
    \ar[r,-,shift right=1pt]
    \ar[d]
    &
    E G
    \ar[
      dl,
      "
        \;\;\;
        \mbox{
          \tiny
          \color{darkblue}
          \bf
          \begin{tabular}{c}
            universal
            \\
            principal
            \\
            $G$-bundle
          \end{tabular}
       }
      "{
         right
      }
    ]
    \\
    \BaseSpace
    \ar[
      rr,
      "
        \vdash P
      ",
      "
        \mbox{
          \tiny
          \color{greenii}
          \bf
          classifying map
        }
      "{
        below
      }
    ]
    &&
    B G
    \,.
  \end{tikzcd}
\end{equation}
Generally, for any (topological/simplicial) action $G \acts \, F$ of $G$ on a
typical fiber $F$, the homotopy quotient serves as the universal
$G$-structured/associated $F$-fiber bundle (\cite[\S 4]{NSS12a}\cite[\S 2.2]{SS20b}):
\begin{equation}
  \label{ClassificationOfAssociatedBundles}
  \begin{tikzcd}[column sep=small]
    E
    \ar[rr]
    \ar[
      d,
      "p"{
        right
      },
      "
        \mbox{
          \tiny
          \color{darkblue}
          \bf
          \begin{tabular}{c}
            $G$-structured
            \\
            $F$-fiber
            bundle
          \end{tabular}
        }
      "{
        left
      }
    ]
    \ar[
      drr,
      phantom,
      "
        \overset{
          \raisebox{1pt}{
            \tiny
            \color{orangeii}
            \bf
            homotopy pullback
          }
        }{
          \mbox{\tiny(pb)}
        }
      "
    ]
    &
    {\phantom{AAAA}}
    &
    F \!\sslash\! G
    \ar[r,-,shift left=1pt]
    \ar[r,-,shift right=1pt]
    \ar[d]
    &
    E G \times_G F
    \ar[
      dl,
      "
       \;\;\;
       \mbox{
         \tiny
         \color{darkblue}
         \bf
         \begin{tabular}{c}
           universal
           \\
           $G$-structured
           \\
           $F$-fiber bundle
         \end{tabular}
       }
      "{
        right
      }
    ]
    \\
    \BaseSpace
    \ar[
      rr,
      "\vdash E"{
        above
      },
      "
        \mbox{
          \tiny
          \color{greenii}
          \bf
          classifying map
        }
      "{
        below
      }
    ]
    &&
    B G
    \,.
  \end{tikzcd}
\end{equation}
Here and in the following, we are notationally suppressing the homotopies
filling all these diagrams.

\medskip

\noindent
{\bf $G$-Structures.}
Throughout,
$n \in \mathbb{N}$ denotes any natural number.
All (fiber-)vector spaces
and, in particular, all (vertical) tangent spaces
are assumed to be finite-dimensional.

\begin{defn}[$G$-Structure]
  \label{GStructure}
 Given a topological group $G$ and a homomorphism
 $\phi : G \xrightarrow{\;} \mathrm{GL}(d)$, we say that

\vspace{-3mm}
\begin{enumerate}[{\bf (i)}]
\setlength\itemsep{-1.5mm}
\item
 {\it $G$-structure on a real vector bundle} $\mathcal{V} \xrightarrow{p} \BaseSpace$
 is a factorization of its classifying map
 $\BaseSpace \xrightarrow{\vdash \mathcal{V}} B \mathrm{O}(d)$
 through $B \phi \colon B H \xrightarrow{\;} B \mathrm{GL}(d)$;

\item  {\it $G$-structure on a real smooth manifold} $M^d$
  is $G$-structure on its tangent vector bundle.
\end{enumerate}
\end{defn}

\begin{remark}[Literature on $G$-structure]
  The notion of $G$-structures as an efficient tool for controlling the
  geometry of super-string compactifications is discussed in \cite{Lott01}\cite{GMPW04}\cite{Koerber10}\cite{Gaillard11}\cite{DDG14}\cite{Prins16} \cite{MinasianPrinsTriendl17}.
  Beware that tradition in Cartan geometry
  insists that the homomorphism $G \xrightarrow{\phi} \mathrm{GL}(d)$
  be injective (e.g. \cite[p. 46]{CapSlovak09}). Since this demand excludes
  common examples of ``$G$-structures''
  like $\mathrm{Spin}$ structures
  (but also $\mathrm{String}$ structures, etc.; and metaplectic structures, etc.)
  without being necessary for the part of the theory that is relevant here, we do not impose it.
  In algebraic topology this more general notion is known as
  {\it $(B G, B \phi)$-structures} (see \cite[\S 1.4]{Kochman96})
  or as
  {\it tangential structures} \cite[Sec. 5]{GMTW06}
  (observing here that
  the canonical inclusion $\mathrm{O}(d) \hookrightarrow \mathrm{GL}(d)$
  is the maximal compact subgroup, so that $B \mathrm{O}(d)  \xrightarrow B \mathrm{GL}(d)$
  is a homotopy equivalence).
  See \cite[\S 4.2]{SS20b} for extensive discussion,
  comparison and further pointers.
\end{remark}

\medskip
\noindent
{\bf Spherical fibrations.}
\begin{defn}[Orthogonal $n$-sphere fiber bundle]
  \label{OrthonalSphereFiberBundle}
  $\,$

\noindent {\bf (i)}   We say that an $S^n$-fiber
  bundle
  $p : X \xrightarrow{\;} \BaseSpace$
  is {\it orthogonal}
  if it is equivalent to unit sphere bundle
  $S(p) : S(\mathcal{V}) \xrightarrow{\;} \BaseSpace$ inside an
  real vector bundle $p : \mathcal{V} \xrightarrow{\;} \BaseSpace$
  (whose structure group may always be taken to be the orthogonal group).

\noindent {\bf (ii)}    This means equivalently that $X$ is associated
  \eqref{ClassificationOfAssociatedBundles} via a classifying map
  $\BaseSpace \xrightarrow{\vdash X} B \mathrm{O}(n+1)$
  to the classifying space for the orthogonal group,
  which fits into a homotopy-pullback diagram of the following form:
  \begin{equation}
    \label{OrthogonalSphereBundleByPullback}
    \begin{tikzcd}[column sep=30pt]
      X
      \ar[rr]
      \ar[
        d,
        "p"{
          right
        },
        "
          \mbox{
            \tiny
            \color{darkblue}
            \bf
            \begin{tabular}{c}
              orthogonal
              \\
              $n$-sphere bundle
            \end{tabular}
          }
        "{
          left
        }
      ]
      \ar[
        drr,
        phantom,
        "
          \overset{
            \raisebox{1pt}{
              \tiny
              \color{orangeii}
              \bf
              homotopy pullback
            }
          }{
            \mbox{\tiny\rm(pb)}
          }
        "
      ]
      &&
      S^n \!\sslash\! \mathrm{O}(n+1)
      \ar[
        d,
        "
          \mbox{
           \tiny
           \color{darkblue}
           \bf
           \begin{tabular}{c}
             universal
             \\
             orthogonal
             \\
             $n$-shere bundle
           \end{tabular}
          }
        "
      ]
      \\
      \BaseSpace
      \ar[
        rr,
        "\vdash X"{
          above
        },
        "
          \mbox{
            \tiny
            \color{greenii}
            \bf
            classifying map
          }
        "{
          below
        }
      ]
      &&
      B \mathrm{O}(n+1)
      \,,
    \end{tikzcd}
  \end{equation}
  where on the top right we have the homotopy quotient
  (the Borel construction)
  of the $n$-sphere $S^n \simeq S(\mathbb{R}^{n+1})$
  by its canonical action of the orthogonal group.

  \noindent {\bf (iii)}   More generally, if a topological group $G$ acts
  continuously on $S^n$, then we say that an {\it $G$-associated} $n$-sphere fiber
  bundle $X \xrightarrow{\;} \BaseSpace$ is one fitting into a homotopy-pullback diagram
  of this form:
  \begin{equation}
    \label{GSphereBundleByPullback}
    \begin{tikzcd}[column sep=30pt]
      X
      \ar[rr]
      \ar[
        d,
        "
          p
        "{
          right
        },
        "
          \mbox{
            \tiny
            \color{darkblue}
            \bf
            \begin{tabular}{c}
              $G$-structured
              \\
               $n$-sphere bundle
            \end{tabular}
          }
        "{
          left
        }
      ]
      \ar[
        drr,
        phantom,
        "
          \overset{
            \raisebox{1pt}{
              \tiny
              \color{orangeii}
              \bf
              homotopy pullback
            }
          }{
            \mbox{\tiny\rm(pb)}
          }
        "
      ]
      &&
      S^n \!\sslash\! G
      \ar[
        d,
        "
          \mbox{
            \tiny
            \color{darkblue}
            \bf
            \begin{tabular}{c}
              universal
              \\
              $G$-structured
              \\
              $n$-sphere bundle
            \end{tabular}
          }
        "
      ]
      \\
      \BaseSpace
      \ar[
        rr,
        "\vdash X"{
          above
        },
        "
          \mbox{
            \tiny
            \color{greenii}
            \bf
            classifying map
          }
        "{
          below
        }
      ]
      &&
      B G
      \,,
    \end{tikzcd}
  \end{equation}
\end{defn}

\begin{example}[Universal orthogonal $n$-sphere fiber bundle]
  \label{UniversalOrthogonaSphereFiberBundle}
Denoting the canonical inclusion of orthogonal groups by
\vspace{-2mm}
\begin{equation}
  \label{InclusionOfOrthogonalGroups}
  \begin{tikzcd}[row sep=-3pt]
    \mathrm{O}(n)
    \ar[
      r,
      hook
    ]
    &
    \mathrm{O}(n+1)
    \\
    A
    \ar[
      r,
      phantom,
      "\longmapsto"
    ]
    &
    \mathrm{diag}(1,A)
    \,,
  \end{tikzcd}
\end{equation}
the universal example of orthogonal $n$-sphere bundles
(Def. \ref{OrthonalSphereFiberBundle}) is equivalent to the
map $B i_n$ of classifying spaces induced from \eqref{InclusionOfOrthogonalGroups}:
\begin{equation}
  \label{UniversalOrthogonalnSphereFibration}
  \begin{tikzcd}[column sep=small]
    S^n
    \ar[
      rr
    ]
    \ar[
      d
    ]
    \ar[
      drr,
      phantom,
      "\mbox{\tiny\rm(pb)}"{
        description
      }
    ]
    &&
    S^n \!\sslash\! \mathrm{O}(n+1)
    \ar[
      d
    ]
    \ar[
      r,
      -,
      shift left=1pt
    ]
    \ar[
      r,
      -,
      shift right=1pt
    ]
    &
    B \mathrm{O}(n)
    \ar[
      dl,
      "
        \;\;\;
        B i_n
      "{
        right
      }
    ]
    \\
    \ast
    \ar[
      rr
    ]
    &&
    B \mathrm{O}(n+1)
  \end{tikzcd}
\end{equation}
\end{example}
\noindent
This example is classical, see for instance \cite[p. 4]{BottCattaneo97}.
But it is just a special case of a more general phenomenon
that will be useful for our purpose:

\begin{example}[Universal $G$-structured $n$-sphere fiber bundle]
  \label{UniversalGStructurednSphereFiberBundle}
  Let
  \vspace{-2mm}
  \begin{equation}
    \label{SubgroupInclusion}
    \begin{tikzcd}
       H
       \ar[
         r,
         hook
       ]
       &
       G
    \end{tikzcd}
  \end{equation}

  \vspace{-2mm}
\noindent  be an inclusion of topological groups.
  Then the homotopy quotient (Borel construction) of their coset space
  $G/H$ by its canonical residual left $G$-action is equivalent to the
  homotopy type of the classifying space of $H$
  (\cite[Lem. 2.7]{FSS19b}):
  \begin{equation}
  \label{UniversalGModHFiberBundle}
  \begin{tikzcd}[column sep=tiny]
    G/H
    \ar[
      rr
    ]
    \ar[
      d
    ]
    \ar[
      drr,
      phantom,
      "\mbox{\tiny\rm(pb)}"{
        description
      }
    ]
    &&
    (G/H) \!\sslash\! G
    \ar[
      d
    ]
    \ar[
      r,
      -,
      shift left=1pt
    ]
    \ar[
      r,
      -,
      shift right=1pt
    ]
    &
    B H
    \ar[
      dl,
      "
        \;\;\;
        B i
      "{
        right
      }
    ]
    \\
    \ast
    \ar[
      rr
    ]
    &&
    B G
  \end{tikzcd}
\end{equation}
Therefore, when the coset space $G/H$ is in fact an $n$-sphere
equipped with $G$-action, which happens
in the following cases (\cite[Rem. 2.9, Prop. 2.23]{FSS19b})
\begin{equation}
\label{CosetSpaceRealizationsOfSpheres}
\mbox{
\def\arraystretch{1.2}
\begin{tabular}{cc}
{\bf Generic} & {\bf Exceptional}
\\
\begin{tabular}{|ccc|c|}
  \hline
  $H$
  &
  \raisebox{-2pt}{
    $\xhookrightarrow{
      \raisebox{-2pt}{\scalebox{.7}{$i$}}
    }$
  }
  &
  $G$
  &
  $G/H$
  \\
  \hline
  \hline
  $\mathrm{O}(n)$
  &
  $\subset$
  &
  $\mathrm{O}(n+1)$
  &
  \multirow{3}{*}{
    $S^n$
  }
  \\
  \cline{1-3}
  $\mathrm{SO}(n)$
  &
  $\subset$
  &
  $\mathrm{SO}(n+1)$
  &
  \\
  \cline{1-3}
  $\mathrm{Spin}(n)$
  &
  $\subset$
  &
  $\mathrm{Spin}(n+1)$
  &
  \\
  \hline
  $\mathrm{SU}(n)$
  &
  $\subset$
  &
  $\mathrm{SU}(n+1)$
  &
  $S^{2n-1}$
  \\
  \hline
  $\mathrm{Sp}(n)$
  &
  $\subset$
  &
  $\mathrm{Sp}(n+1)$
  &
  $S^{4n-1}$
  \\
  \hline
\end{tabular}
&
\hspace{.3cm}
\begin{tabular}{|ccc|c|}
  \hline
  $H$
  &
  \raisebox{-2pt}{
    $\xhookrightarrow{
      \raisebox{-2pt}{\scalebox{.7}{$i$}}
    }$
  }
  &
  $G$
  &
  $G/H$
  \\
  \hline
  \hline
  $\mathrm{Sp}(1) \times \mathrm{Sp}(1)$
  &
  $\subset$
  &
  $\mathrm{Sp}(2)$
  &
  $S^4$
  \\
  \hline
  $\mathrm{SU}(3)$
  &$\subset$&
  $G_2$
  &
  $S^6$
  \\
  \hline
  $G_2$
  &
  $\subset$
  &
  $\mathrm{Spin}(7)$
  &
  $S^7$
  \\
  \hline
  $\mathrm{Spin}(7)$
  &
  $\subset$
  &
  $\mathrm{Spin}(9)$
  &
  $S^{15}$
  \\
  \hline
  \multicolumn{4}{c}{
    {\phantom{A}}
  }
\end{tabular}
\end{tabular}
}
\end{equation}
then the universal $G$-associated $n$-sphere bundle
is equivalently the classifying space of $H$ (\cite[Prop. 2.8]{FSS19b}):
\begin{equation}
  \label{UniversalGAssociatedSphereBundle}
  \begin{tikzcd}[column sep=tiny]
    S^n
    \ar[
      rr
    ]
    \ar[
      d
    ]
    \ar[
      drr,
      phantom,
      "\mbox{\tiny\rm(pb)}"{
        description
      }
    ]
    &&
    S^n \!\sslash\! G
    \ar[
      d
    ]
    \ar[
      r,
      -,
      shift left=1pt
    ]
    \ar[
      r,
      -,
      shift right=1pt
    ]
    &
    B H
    \ar[
      dl,
      "
        \;\;\;
        B i
      "{
        right
      }
    ]
    \\
    \ast
    \ar[
      rr
    ]
    &&
    B G
  \end{tikzcd}
\end{equation}
\end{example}

\medskip

\noindent
{\bf Vertical tangent bundles to spherical fibrations.}
We now show how this universal homotopy-theoretic construction
of sphere bundles
knows everything about their vertical tangent bundles.
\begin{prop}[Classifying map of frame bundle to $n$-sphere]
  \label{ClassifyingMapToFrameBundleTonSphere}
  Under the identification on the right of
  \eqref{UniversalOrthogonalnSphereFibration},
  the homotopy fiber inclusion $\mathrm{fib}(B i_n)$ of $S^n$
  into the universal orthogonal $n$-sphere fiber bundle
  (Example \ref{UniversalOrthogonaSphereFiberBundle})
  is the classifying map
  $\vdash \mathrm{Fr}(S^n)$ for the orthogonal frame bundle
  $\mathrm{Fr}_{\mathrm{O}}(S^n) \to S^n$ of the $n$-sphere:
  \begin{equation}
  \begin{tikzcd}[column sep=35pt]
    S^n
    \ar[
      rr,
      "
        \vdash \mathrm{Fr}_{\mathrm{O}}(S^n)
      "{
        below
      },
      "
        \mbox{
         \tiny
         \color{greenii}
         \bf
         \begin{tabular}{c}
           classifying map of
           \\
           orthogonal frame bundle
         \end{tabular}
        }
      "{
        above
      }
    ]
    \ar[
      d
    ]
    \ar[
      drr,
      phantom,
      "
        \raisebox{-10pt}{
          \mbox{\tiny\rm(pb)}
        }
      "
    ]
    &&
    B \mathrm{O}(n)
    \ar[
      d,
      "B i_n"
    ]
    \\
    \ast
    \ar[
      rr
    ]
    &&
    B \mathrm{O}(n+1)
  \end{tikzcd}
  \end{equation}
\end{prop}
\begin{proof}
The long homotopy fiber sequence of $B i_n$
(e.g. \cite[Def. 2.26]{NSS12a}, following from the pasting law \cite[Prop. 2.23]{NSS12a})
shows that the homotopy fiber inclusion of
$B i_n$ classifies an $\mathrm{O}(n)$-principal bundle over
the $n$-sphere whose total space is $\mathrm{O}(n+1)$
with $\mathrm{O}(n)$-action induced by the
canonical inclusion $i_n$ \eqref{InclusionOfOrthogonalGroups}:
$$
\begin{tikzcd}
  \mathrm{O}(n)
  \ar[
    r,
    hook,
    "i"
  ]
  \ar[d]
  \ar[
    dr,
    phantom,
    "\mbox{\tiny\rm(pb)}"{
      description
    }
  ]
  &
  \mathrm{O}(n+1)
  \ar[
    r
  ]
  \ar[
    d
  ]
  \ar[
    dr,
    phantom,
    "\mbox{\tiny\rm(pb)}"{
      description
    }
  ]
  &
  \ast
  \ar[
    d
  ]
  \\
  \ast
  \ar[r]
  &
  S^n
  \ar[
    r,
    "\mathrm{fib}(B i_n)"{
      description
    }
  ]
  \ar[d]
  \ar[
    dr,
    phantom,
    "\mbox{\tiny\rm(pb)}"{
      description
    }
  ]
  &
  B \mathrm{O}(n)
  \ar[
    d,
    "B i_n"
  ]
  \\
  &
  \ast
  \ar[
    r
  ]
  &
  B \mathrm{O}(n+1)
\end{tikzcd}
$$
Therefore, it is sufficient to observe that
we have an isomorphism of $\mathrm{O}(n)$-principal bundles
\vspace{-2mm}
$$
\begin{tikzcd}[column sep=small, row sep=-2pt]
  \mathrm{O}(n+1)
  \ar[
    out=180-66,
    in=66,
    looseness=3.5,
    "
    \scalebox{.77}{$
      \phantom{\cdot}
      \mathclap{
        \mathrm{O}(n)
      }
      \phantom{\cdot}
    $}
    "{
      description
    },
    shift right=1
  ]
  \ar[
    rr,
    "\sim"
  ]
  &&
  \mathrm{Fr}(S^n)
  \ar[
    out=180-66,
    in=66,
    looseness=3.5,
    "
    \scalebox{.77}{$
      \phantom{\cdot}
      \mathclap{
        \mathrm{O}(n)
      }
      \phantom{\cdot}
    $}
    "{
      description
    },
    shift right=1
  ]
  \\
  A
  \ar[
    rr,
    phantom,
    "\mapsto"{
      description
    }
  ]
  &&
  \big(
    A \cdot v_1,
    \cdots,
    A \cdot v_n
  \big)
  \mathrlap{
    \in
    \mathrm{Fr}_{A \cdot v_0}
    (S^n)
    \,,
  }
  \,
\end{tikzcd}
$$

\vspace{-3mm}
\noindent where $v_0, \cdots v_n \in \mathbb{R}^{n+1}$
are the canonical basis vectors and where on the right
we regard $S^n = S(\mathbb{R}^{n+1})$ with the induced
identification of
$T_{A \cdot v_0} S^n \,\simeq\, (A \cdot v^0)^\perp \subset \mathbb{R}^{n+1}$.
\hfill \end{proof}

More generally:
\begin{prop}[Classifying map of $H$-frame bundle of $H$-coset realization of $n$-sphere]
  \label{ClassifyingMapOfHFrameBundleTonSphere}
  Given a coset-space realization of the $n$-sphere
  $S^n \simeq G/H$ \eqref{CosetSpaceRealizationsOfSpheres}
  induced from a subgroup inclusion $H \xrightarrow{i} G$ \eqref{SubgroupInclusion}
  of compact Lie groups,
  then under the identification on the right of \eqref{UniversalGAssociatedSphereBundle}
  the homotopy fiber inclusion $\mathrm{fib}(B i)$ of $S^n$ into the universal
  $G$-associated $n$-sphere fiber bundle (Example \ref{UniversalGStructurednSphereFiberBundle})
  is a classifying map for
  the $H$-principal bundle on the $n$-sphere which exhibits
  its canonical $H$-structure (Def. \ref{GStructure}):
  \vspace{-2mm}
  \begin{equation}
  \begin{tikzcd}[column sep=35pt]
    S^n
    \ar[
      rr,
      "
        \vdash \mathrm{Fr}_H(S^n)
      "{
        below
      },
      "
        \mbox{
         \tiny
         \color{greenii}
         \bf
         \begin{tabular}{c}
           classifying map of
           \\
           $H$-frame bundle
         \end{tabular}
        }
      "{
        above
      }
    ]
    \ar[
      d
    ]
    \ar[
      drr,
      phantom,
      "
        \raisebox{-10pt}{
          \mbox{\tiny\rm(pb)}
        }
      "
    ]
    &&
    B H
    \ar[
      d,
      "B i"
    ]
    \\
    \ast
    \ar[
      rr
    ]
    &&
    B G
  \end{tikzcd}
  \end{equation}
\end{prop}
\begin{proof}
As before, the long homotopy fiber sequence of $B i$
(e.g. \cite[Def. 2.26]{NSS12a}, following from the pasting law \cite[Prop. 2.23]{NSS12a})
shows that the homotopy fiber inclusion of
$B i$ classifies an $H$-principal bundle over
the $n$-sphere whose total space is $G$
with $H$-action induced by the
given subgroup inclusion:
$$
\begin{tikzcd}[column sep=large]
  H
  \ar[
    r,
    hook,
    "i"
  ]
  \ar[d]
  \ar[
    dr,
    phantom,
    "\mbox{\tiny\rm(pb)}"{
      description
    }
  ]
  &
  G
  \ar[
    r
  ]
  \ar[
    d
  ]
  \ar[
    dr,
    phantom,
    "\mbox{\tiny\rm(pb)}"{
      description
    }
  ]
  &
  \ast
  \ar[
    d
  ]
  \\
  \ast
  \ar[r]
  &
  S^n
  \ar[
    r,
    "\mathrm{fib}(B i)"{
      description
    }
  ]
  \ar[d]
  \ar[
    dr,
    phantom,
    "\mbox{\tiny\rm(pb)}"{
      description
    }
  ]
  &
  B H
  \ar[
    d,
    "B i"
  ]
  \\
  &
  \ast
  \ar[
    r
  ]
  &
  B G
\end{tikzcd}
$$
Therefore, it is sufficient to observe that $G \xrightarrow{\;} G/H$ is
an $H$-frame bundle that exhibits $H$-structure (Def. \ref{GStructure})
on the tangent bundle $T(G/H)$.
This is basic fact of Cartan geometry, laid out for instance in \cite[p. 53]{CapSlovak09}.
\hfill \end{proof}

\begin{example}[Canonical Spin structure on $n$-spheres]
  For the generic coset space realization of the $n$-sphere from \eqref{CosetSpaceRealizationsOfSpheres},
  $S^n \simeq \mathrm{Spin}(n+1)/\mathrm{Spin}(n)$,
  Prop. \ref{ClassifyingMapOfHFrameBundleTonSphere} says that the homotopy fiber
  inclusion of the map of classifying spaces
  $B \mathrm{Spin}(n) \xrightarrow B \mathrm{Spin}(n+1)$ classifies
  a $\mathrm{Spin}(n)$-principal bundle
  of the form  $\mathrm{Spin}(n+1) \xrightarrow{\;} S^n$
  and that this is a Spin structure (Def. \ref{GStructure}) on the $n$-sphere.
  A traditional proof of this fact is spelled out
  in detail in \cite[Thm. A.6.6]{Nowaczyk15}, see also \cite[\S 2.a]{Gutt88}.
\end{example}

\begin{prop}[$H$-Structure on vertical tangent bundle of $G$-associated sphere bundle]
  \label{HStructureOnVerticalTangentBundleOfGAssociatedSphereBundle}
  Given a coset-space realization of the $n$-sphere
  $S^n \simeq G/H$ \eqref{CosetSpaceRealizationsOfSpheres}
  induced from a Lie subgroup inclusion $H \xrightarrow{i} G$ \eqref{SubgroupInclusion},
  then for a
  $G$-associated $S^n$-fiber bundle $X \xrightarrow{S(p)} \BaseSpace$
  \eqref{GSphereBundleByPullback}:

\vspace{-3mm}
\begin{enumerate}[{\bf (i)}]
\setlength\itemsep{-.08cm}
 \item  the vertical tangent bundle carries an $H$-structure (Def. \ref{GStructure});

  \item
  whose associated $G$-principal bundle is the pullback along $S(p)$
  of that to which $X$ is associated.
\end{enumerate}
\end{prop}
\begin{proof}
By the classification \eqref{ClassificationOfAssociatedBundles}
of fiber bundles, $S(p)$ sits in a homotopy pullback square
as on the right of the following pasting diagram
\begin{equation}
  \label{CharacterizationOfVerticalTangentBundlesOfSphereBundles}
  \begin{tikzcd}[column sep=tiny]
    S^n_q
    \ar[
      rr
    ]
    \ar[
      rrrrr,
      bend left=20,
      "
        \vdash T_{S(p_q)} X
      "
    ]
    \ar[d]
    \ar[
      drr,
      phantom,
      "\mbox{\tiny\rm(pb)}"
    ]
    &&
    X
    \ar[
      d,
      "S(p)"{
        description
      }
    ]
    \ar[
      rr,
      "\vdash T_{S(p)} X"{
        description
      }
    ]
    \ar[
      drr,
      phantom,
      "\mbox{\tiny\rm(pb)}"
    ]
    &
    {\phantom{AAAA}}
    &
    S^n \!\sslash\! G
    \ar[r,-,shift left=1pt]
    \ar[r,-,shift right=1pt]
    \ar[d]
    &
    B H
    \ar[
      dl
    ]
    \\
    \ast
    \ar[
      rr,
      "x"
    ]
    &
    {\phantom{AAA}}
    &
    \BaseSpace
    \ar[
      rr,
      "\vdash X"{
        above
      }
    ]
    &&
    B G
    \,.
  \end{tikzcd}
\end{equation}
  Notice that if the top map
  in the right square classifies $H$-structure on
  the vertical tangent bundle, as indicated by its label,
  then the homotopy-commutativity of the right square is
  equivalent to claim (ii).

  Hence it is sufficient now to prove that the
  top right map indeed classifies $H$-structure on the vertical tangent bundle;
  which then also proves claim (i).

  To that end, consider any point $q \in \BaseSpace$ and write
  $S^n_q$ for the sphere fiber over it, as shown by the
  homotopy pullback square on the left of \eqref{CharacterizationOfVerticalTangentBundlesOfSphereBundles}.
  By the
  pasting law (\cite[Prop. 2.23]{NSS12a}) it follows that the
  full rectangle is a homotopy pullback.
  Therefore Prop. \ref{ClassifyingMapOfHFrameBundleTonSphere}
  says that the composite top map in \eqref{CharacterizationOfVerticalTangentBundlesOfSphereBundles}
  classifies $H$-structure on the tangent bundle of $S^n_q$.
  Since this true for all $q$, it follows that
  the $H$-principal bundle classified by
  $\vdash T_{S(p)} X$ restricts on each sphere fiber $S^n_q$
  to that sphere's tangent $H$-structure. But this is the defining property of
  ($H$-structure on) the vertical tangent bundle of $X$.
\hfill \end{proof}

\begin{cor}[Once-stabilized vertical tangent bundle of orthogonal sphere bundle is basic]
  \label{OnceStabilizedVerticalTangentBundleOfSphereBundleIsBasic}
  The once-stabilized vertical tangent bundle
  to an orthogonal sphere bundle $S(p) \colon S(\mathcal{V}) \to Q$
  \eqref{OrthogonalSphereBundleByPullback}
  is isomorphic to the pullback of its underlying
  vector bundle:
  \begin{equation}
    \label{VerticalTangentBundleToOrthogonalSphereBundle}
    \overset{
      \mathclap{
      \raisebox{3pt}{
        \tiny
        \color{darkblue}
        \bf
        \begin{tabular}{c}
          vertical
          \\
          tangent bundle
        \end{tabular}
      }
      }
    }{
      \overbrace{
        T_{S(p)}
        \big(
          \underset{
            \mathclap{
            \raisebox{-3pt}{
              \tiny
              \color{darkblue}
              \bf
              \begin{tabular}{c}
                orthogonal
                \\
                sphere bundle
              \end{tabular}
            }
            }
          }{
          \underbrace{
            S(\mathcal{V})
          }
          }
        \big)
      }
    }
    \overset{
      \raisebox{3pt}{
        \tiny
        \color{darkblue}
        \bf
        \begin{tabular}{c}
          one-step
          \\
          stabiliz.
        \end{tabular}
      }
    }{
      \overbrace{
        \times
        \mathbb{R}
      }
    }
    \;\;
      \simeq_{{}_{\BaseSpace}}
    \;\;
    \overset{
     \mathclap{
      \raisebox{3pt}{
       \tiny
       \color{darkblue}
       \bf
       \begin{tabular}{c}
         pullback of
         \\
         associated vector bundle
       \end{tabular}
      }
      }
    }{
    \overbrace{
      S(p)^\ast
      \big(
        \mathcal{V}
      \big)
    }
    }
    \,.
  \end{equation}
\end{cor}
\noindent
A traditional proof of this statement is indicated in \cite[Prop. 1.1.9]{Gollinger16}.
\begin{proof}
  For the
  orthogonal subgroup inclusion
  $\mathrm{O}(n) \xrightarrow{i_n} \mathrm{O}(n+1)$
  \eqref{InclusionOfOrthogonalGroups},
  Prop. \ref{HStructureOnVerticalTangentBundleOfGAssociatedSphereBundle}
  gives a homotopy pullback diagram
  \eqref{VerticalTangentBundleToOrthogonalSphereBundle}
  of this form:
\begin{equation}
  \begin{tikzcd}[column sep=small]
    S(\mathcal{V})
    \ar[
      rrr,
      bend left=15pt,
      "\vdash T_{S(p)} S(\mathcal{V})"
    ]
    \ar[
      rr
    ]
    \ar[
      d,
      "
        S(p)
      "{
        left
      }
    ]
    \ar[
      drr,
      phantom,
      "\mbox{\tiny\rm(pb)}"
    ]
    &
    {\phantom{AA}}
    &
    S^n /\!\!/ \mathrm{O}(n+1)
    \ar[
      r,
      -,
      shift left=1pt
    ]
    \ar[
      r,
      -,
      shift right=1pt
    ]
    \ar[d]
    &
    B \mathrm{O}(n)
    \ar[
      dl,
      "
        \;\;\;
        B i_n
      "{
        right
      }
    ]
    \\
    \BaseSpace
    \ar[
      rr,
      "
        \vdash \mathcal{V}
      "{
        above
      }
    ]
    &&
    B \mathrm{O}(n+1)
  \end{tikzcd}
\end{equation}
Noticing that postcomposition with $B i_{n}$ manifestly corresponds to
one-step stabilization of an orthogonal vector bundle,
the homotopy-commutativity of this square is exactly the claim to be proven.
\hfill \end{proof}

The following is stated without proof as \cite[Fact 3.1]{CrowleyEscher03},
apparently reading between the lines of \cite[p. 403]{Milnor56}.

\begin{cor}[Once-stabilized tangent bundle of orthogonal sphere bundle is basic]
  \label{OnceStabilizedTangentBundleOfSphereBundleIsPulledBackFromBase}
  $\,$

  \noindent
  If the base space $\BaseSpace$ is a smooth manifold,
  then the once-stabilized tangent bundle of the
  total space of an orthogonal sphere bundle $S(p) :  S(\mathcal{V}) \xrightarrow{\;} X$
  (Def. \ref{OrthonalSphereFiberBundle})
  is isomorphic to the
  pullback along $S(p)$ of the Whitney sum of the tangent bundle of
  the base with the given vector bundle:
  \begin{equation}
    T \big( S(\mathcal{V})\big)
    \times \mathbb{R}
    \;\simeq_{{}_{\BaseSpace}}\;
    S(p)^\ast
    \big(
      T \BaseSpace \oplus_{{}_\BaseSpace} \mathcal{V}
    \big).
  \end{equation}
\end{cor}

\begin{proof}
  Consider the following sequence of bundle isomorphisms over the base space
  $\BaseSpace$:
  $$
    \begin{aligned}
      T
      \big(
        S(\mathcal{V})
      \big)
      \times
      \mathbb{R}
      \;
      &
      \simeq_{\BaseSpace}
      \big(
        S(p)^\ast
        (
          T \BaseSpace
        )
      \big)
      \oplus_{{}_Q}
      \Big(
        T_{S(p)}
        \big(
          S(\mathcal{V})
        \big)
        \times
        \mathbb{R}
      \Big)
      \\
      & \simeq_{{}_Q}
      \big(
        S(p)^\ast
        (
          T \BaseSpace
        )
      \big)
      \oplus_{{}_Q}
      \big(
        S(p)^\ast
        \mathcal{V}
      \big)
      \\
      & \simeq_{{}_Q}
      S(p)^\ast
      \big(
        T \BaseSpace
        \oplus_{{}_Q}
        \mathcal{V}
      \big)
      \,.
    \end{aligned}
  $$

  \vspace{-2mm}
  \noindent Here:

  \vspace{-.2cm}
  \begin{enumerate}

  \vspace{-.2cm}
  \item[(a)]
  The first step is a splitting of the short exact sequence of
  vector bundles
  \vspace{-2mm}
  $$
    \begin{tikzcd}[column sep=small]
      0
      \ar[r]
      &
      T_{S(p)} S(\mathcal{V})
      \ar[
        rr,
        hook
      ]
      &&
      T
      \big(
        S(\mathcal{V})
      \big)
      \ar[
        rr,
        ->>,
        "d S(p)"
      ]
      &&
      S(p)^\ast
      (T \BaseSpace)
      \ar[r]
      &
      0
    \end{tikzcd}
  $$

    \vspace{-4mm}
\noindent   that defines the vertical tangent bundle $T_{S(p)} S(\mathcal{V})$,
  and which splits as a special case of the general splitting of
  short exact sequences of real vector bundles over paracompact Hausdorff base spaces,
  in particular over smooth manifolds, 
  by forming orthogonal complements with respect to any choice of a continuous fiberwise inner product.

  \vspace{-.2cm}
  \item[(b)]
  The second step is Corollary \ref{OnceStabilizedVerticalTangentBundleOfSphereBundleIsBasic}.

  \vspace{-.2cm}
  \item[(c)]
  The last step is the distributivity of pullback over
  Whitney sum of vector bundles.
  \end{enumerate}

\vspace{-8mm}
\hfill \end{proof}

\noindent
In conclusion :
\begin{prop}[Stable characteristic classes on sphere bundles are basic]
  \label{StableCharacteristicClassesOnSphereBundlesAreBasic}
  Given an orthogonal sphere-fiber bundle $S(\mathcal{V})$,

 \noindent {\bf (i)}  every {\it stable} characteristic class
  -- hence every polynomial $P(p_1, p_2, \cdots)$ of
  Pontrjagin classes $p_i$ --
  of its vertical tangent bundle is basic,
  i.e.: pulled back from the base
  space $\BaseSpace$:
  $$
    P(p_1, p_2, \cdots)
    \big(
      T_{S(p)} S(\mathcal{V})
    \big)
    \;=\;
    S(p)^\ast
    \underset{
      \in
      H^\bullet
      (
        \BaseSpace
      )
    }{
    \underbrace{
      P(p_1, p_2, \cdots)
      \big(
        \mathcal{V}
      \big)
    }
    }
    \,;
  $$

  \vspace{-2mm}
\noindent {\bf (ii)}   and if the base space $\BaseSpace$ is
  a smooth manifold then then analogous statement holds for every stable class
  of the full tangent bundle
  \vspace{-2mm}
  $$
    P(p_1, p_2, \cdots)
    \big(
      T S(\mathcal{V})
    \big)
    \;=\;
    S(p)^\ast
    \underset{
      \in
      H^\bullet
      (
        \BaseSpace
      )
    }{
    \underbrace{
      P(p_1, p_2, \cdots)
      (
        T\BaseSpace \oplus_{{}_{Q}} \mathcal{V}
      )
    }
    }
    \,;
  $$
\end{prop}
\begin{proof}
  Since a stable characteristic class of a vector bundle, such as a Pontrjagin class,
  is one that can be pulled back from any direct sum of that vector bundle
  with a trivial vector bundle,
  the first claim follows
  by
  Cor. \ref{OnceStabilizedVerticalTangentBundleOfSphereBundleIsBasic}
  and the second by
  Cor. \ref{OnceStabilizedTangentBundleOfSphereBundleIsPulledBackFromBase}.
\hfill \end{proof}


\medskip

\vspace{1cm}
\noindent  Hisham Sati, {\it Mathematics, Division of Science, New York University Abu Dhabi, UAE.}
\\
{\tt hsati@nyu.edu}
\\
\\
\noindent  Urs Schreiber, {\it Mathematics, Division of Science, New York University Abu Dhabi, UAE;
on leave from Czech Academy of Science, Prague.}
\\
{\tt us13@nyu.edu}

\end{document}